\documentclass[aps,prl,showpacs,floats,twocolumn,floats,superscriptaddress,floatfix]{revtex4-1}
\usepackage{graphicx}
\usepackage{bm}
\usepackage{amsfonts}
\usepackage{color}
\usepackage{amsmath}    % need for subequations
\usepackage{epsfig}
\usepackage{subfigure}  % use for side-by-side figures
\usepackage{hyperref}   % use for hypertext links, including those to external documents and URLs
\usepackage{bm}
\usepackage{amssymb}
\usepackage{soul}
\usepackage{xcolor}
\usepackage{float}

\begin{document}
\title{The geometry of planar linear flows}
\author{Sabarish V Narayanan}
\email{sabarish@jncasr.ac.in}
\affiliation{Jawaharlal Nehru Centre for Advanced Scientific Research, Bangalore 560064, India}
\author{Ganesh Subramanian}
 \email{sganesh@jncasr.ac.in}
\affiliation{Jawaharlal Nehru Centre for Advanced Scientific Research, Bangalore 560064, India}

\begin{abstract}

We identify incompressible planar linear flows that are generalizations of the well known one-parameter family characterized by the ratio of in-plane extension to (out-of-plane) vorticity. The latter \textquoteleft canonical' family is classified into elliptic and hyperbolic linear flows with closed and open streamlines, respectively, corresponding to the extension-to-vorticity ratio being less or greater than unity; unity being the marginal case of simple shear flow. The novel flows possess an out-of-plane extension, but the streamlines may nevertheless be closed or open, allowing for an organization, in a three-dimensional parameter space, into regions of \textquoteleft eccentric' elliptic and hyperbolic flows, separated by a surface of degenerate linear flows with parabolic streamlines that are generalizations of simple shear. We discuss implications for various fluid mechanical scenarios.

\end{abstract}

% \pacs{47.27.Gs, 05.20.Jj}
\maketitle  

The one-parameter planar linear flow family is an important class of incompressible flows illustrating the fluid mechanical consequences of varying the relative magnitudes of extension and vorticity\cite{LealBook,Kao,Bentley,Marath2}. Limiting members include planar extension (zero vorticity) and solid body rotation (zero extension) with simple shear, the transitional member with straight streamlines, separating hyperbolic and elliptic flows with open and closed streamlines, respectively. The planar linear flow family is of particular importance in microhydrodynamics; the microstructural element of a sheared complex fluid sees a local linear flow, and for an ambient two-dimensional flow, this must be one of the planar linear flows. Simple shear is rheologically significant, being the flow topology generated in standard rheometric devices. Drop deformation and break-up in planar linear flows have been extensively studied\cite{Taylor,Bentley2,Bentley, RallisonRev, StoneRev}; recent efforts focus on vesicles\cite{dechampssteinberg2009,ZabuskySteinberg2011,vivek2019} and the hysteretic dynamics of macromolecules in these flows\cite{Larson, Shaqfeh1, Chu, Shaqfeh2, HurShaqfeh2002, ChuShaqfeh2003, HoffmanShaqfeh2007}. Planar linear flows highlight diffusion-limited transport from suspended particles and drops in the Stokesian regime\cite{Frankel1, Acrivos, Poe,Deepak1}, thereby emphasizing the role of streamline topology in convective enhancement\cite{Sub1, Sub2, Deepak1, Deepak2, Mahan}. These flows are crucial to dilute suspension rheology -\! closed pathlines in the two-sphere problem\cite{Batchelor,Kao} and closed\,(Jeffery) orbits of an axisymmetric particle\cite{Jeffery,Bretherton,Hinch1,Hinch2}, in subsets of these flows, lead to an indeterminate stress in the pure-hydrodynamic limit\cite{Batchelor2,Brady, Bergenholtz}, rendering the rheology sensitive to departures from Stokesian hydrodynamics\cite{Batchelor2,Bergenholtz,SubBrady2006,Hinch1,Marath1, Marath2,Marath3}. The above examples showcase the importance of what, from hereon, is termed the canonical planar linear flow family. It forms an integral element of any flow-classification scheme based on microstructural evolution\cite{Olbricht}.
%\begin{figure*}[]
%\centering
%\includegraphics[height = 0.65\columnwidth, width=2\columnwidth]{Fig_0_new.png}\\
%\includegraphics[scale=0.35]{Fig_0_new.png}
%\vspace*{-0.15in}\caption{The $Q$-axis, along with the surface\,(red) and 3D\,(blue/black) streamline patterns for the canonical planar linear flows; $\bm{\omega}$ is the vorticity vector, and $E_{1,2}$ are the principal in-plane extensions\,($E_3=0$).}\vspace*{-0.3in}
%\label{fig:1}
%\end{figure*}

\begin{figure}[h]
\includegraphics[width = 1.075\columnwidth, height = 0.725\columnwidth]{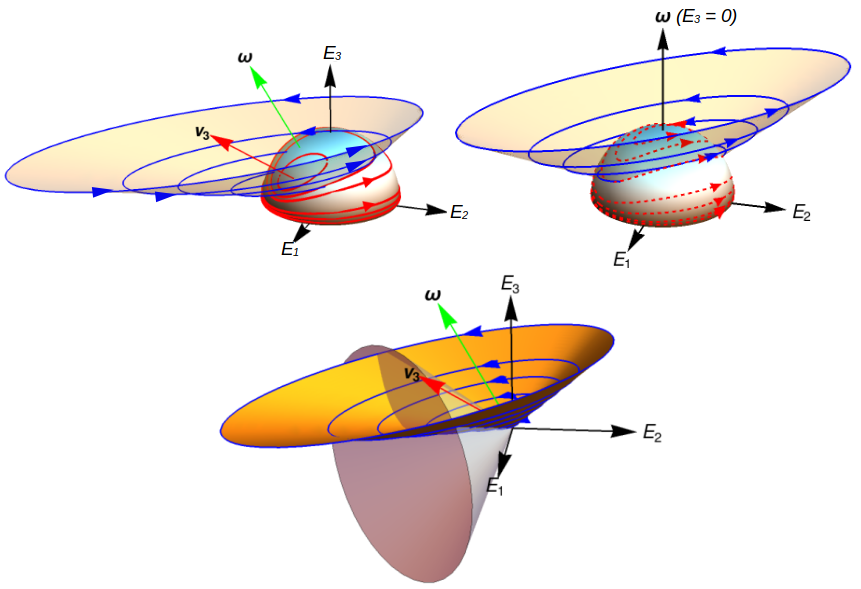}
\vspace*{-0.15in}\caption{Surface\,(red) and 3D\,(blue) streamlines of (a) eccentric\,($\alpha\!=\!4\sqrt{2}, \epsilon\!=\!-2,\theta_\omega\!=\!\pi/3,\phi_\omega\!=\!0$) and (b) canonical\,($\hat{\alpha} =- 0.5$) elliptic linear flows. (c) The neutral and inclined elliptical cones for the eccentric case; ${\boldsymbol v}_3$ lies on the neutral (circular)cone, and is the axis of the inclined elliptical cone of streamlines; $E_i(i\!=\!1\!-\!3)$ are the principal extensions.}\vspace*{-0.3in}
\label{fig:2}
\end{figure}

A linear flow is defined by $\!\bm{u}\!=\! \bm{\Gamma}\!\bm{\cdot}\!\bm{x}$, $\bm{\Gamma}$ being the transpose of the velocity gradient tensor. $\bm{\Gamma} \!\!=\!\!\left[\!\begin{smallmatrix} 0 & 1 & 0 &\\ \hat{\alpha} & 0 & 0 \\ 0 & 0 & 0\!\end{smallmatrix}\!\right]$ for canonical planar linear flows with $\hat{\alpha}\!\! \in\!\! [-1, 1]$;\! $\frac{(1+\hat{\alpha})}{(1-\hat{\alpha})}$ denotes the ratio of in-plane extension to (out-of-plane) vorticity. The eigenvalues of $\bm{\Gamma}$, $\mu_i\! \!\equiv\!\! \!\{ -\sqrt{\hat{\alpha}}, \sqrt{\hat{\alpha}}, 0\}$, imply a trivial cubic invariant\,($R \!= \!\mu_1 \mu_2 \mu_3\! = \!0$), with the quadratic invariant $Q\!\! =\!\! \mu_1 \mu_2\! +\! \mu_2 \mu_3\! +\! \mu_1 \mu_3 \!\!=\! \!-\hat{\alpha}$ determining the nature of the flow. In the $QR$-classification of incompressible linear flows\cite{Perry,Cantwell}, canonical planar linear flows lie on the $Q$-axis:$Q\!\!>\!\!0$ and $Q\!<\!0$ correspond to canonical elliptic and hyperbolic linear flows, respectively, with the streamlines being planar ellipses and hyperbolae; $Q\!=\!0$ corresponds to simple shear. Incompressible linear flows are uniquely characterized by their surface streamline topologies, obtained by projecting the 3D streamline pattern onto a unit sphere. As shown in Fig.\,\ref{fig:2}b, surface streamlines for canonical elliptic flows are spherical ellipses or Jeffery orbits\cite{Jeffery,Bretherton}, with the geometrical aspect ratio replaced by an effective one\,($\!1/\!\sqrt{-\hat{\alpha}}$) in terms of the flow-type parameter. Those for hyperbolic flows are open, and may be regarded as orbits with imaginary aspect ratios\cite{Deepak1}; simple shear yields a meridional surface-streamline topology. The origin,  a fixed point in the dynamical systems parlance\cite{Perry}, is a center for canonical elliptic flows and a saddle for hyperbolic ones, with eigenvectors:$\bm{v_1}\!=\! \{-(\sqrt{\hat{\alpha}})^{-1}, 1, 0\}$; \!\!$\bm{v_2}\!=\!\{(\sqrt{\hat{\alpha}})^{-1},1,0\}$; \!\!$\bm{v_3}\!=\!\{0,0,1\}$. The vorticity $\bm{\omega}$ is normal to the flow plane, and aligned with the neutral eigenvector\,($\bm{v_3}$); see \cite{suppmaterial}.
%\begin{figure}[h]
%\includegraphics[width = \columnwidth, height = 0.7\columnwidth]{Fig_1_new.png}
%\vspace*{-0.15in}\caption{Surface\,(red) and 3D\,(blue) streamlines of (a) eccentric\,($\alpha\!=\!4\sqrt{2}, \epsilon\!=\!-2,\theta_\omega\!=\!\pi/3,\phi_\omega\!=\!0$) and (b) the neutral and inclined elliptical cones for the eccentric case; ${\boldsymbol v}_3$ lies on the neutral (circular)cone, and is the axis of the inclined elliptical cone with 3D streamlines mapping onto a single surface streamline; $E_i(i\!=\!1\!-\!3)$ are the principal extensions.}\vspace*{-0.3in}
%\label{fig:2}
%\end{figure}

We report generalizations of the canonical planar linear flows above, where streamlines are plane (quadratic) curves despite a non-trivial extensional component normal to the flow plane. The existence of these novel flows is most easily understood for the elliptic case, the surface streamline topology for which is shown in Fig.\,\ref{fig:2}a; despite again being spherical ellipses, the eccentric disposition of the surface streamlines is in contrast to the concentric arrangement in the canonical case(Fig.\,\ref{fig:2}b). The limiting surface streamlines in both cases comprise a great circle and the point in which $\bm{v_3}$ intersects the unit sphere. But, $\bm{v_3}$ for the eccentric case is neither aligned with ${\bm{\omega}}$, nor orthogonal to the plane of the flow. 

The `eccentric' elliptic linear flow\,(Fig.\,\ref{fig:2}a) may be rationalized by first noting that a $3D$ extensional flow has an associated cone of neutral directions\,(${\boldsymbol n}$), defined by ${\boldsymbol \Gamma}\!\!:\!\!{\boldsymbol n}{\boldsymbol n}\!=\!0$. The cone, a circular one for axisymmetric extension and an elliptical one in general, separates regions with positive and negative rates of stretch; its interior corresponds to a positive rate of stretch when the largest\,(in magnitude) eigenvalue is positive. For $\boldsymbol \omega\!=\!0$, any material point comes from, and goes to, infinity, crossing the neutral cone in the process. But, for an appropriately oriented ${\boldsymbol \omega}$ of the right magnitude, a material point traverses a closed curve, since the line element connecting this point to the origin samples regions within and outside the cone such that the net rate of stretch over a $2\pi$-circuit is zero. Linearity of the flow implies that this curve is an ellipse, and that all streamlines are geometrically similar ellipses\,(the blue curves in Fig.\ref{fig:2}a); ${\boldsymbol v}_3$ lies on the neutral cone, the line along ${\boldsymbol v}_3$ being the locus of ellipse centers(Fig.\ref{fig:2}c). Each surface streamline is the projection of closed streamlines on an inclined elliptical cone with axis along ${\boldsymbol v}_3$. Elliptical double cones, with the vertex solid angle ranging from zero\,(the straight line along ${\boldsymbol v}_3$) to $2\pi$\,(the plane of the flow), foliate 3D space.

\begin{figure}[h]
\includegraphics[width = 1\columnwidth, height = 0.475\columnwidth]{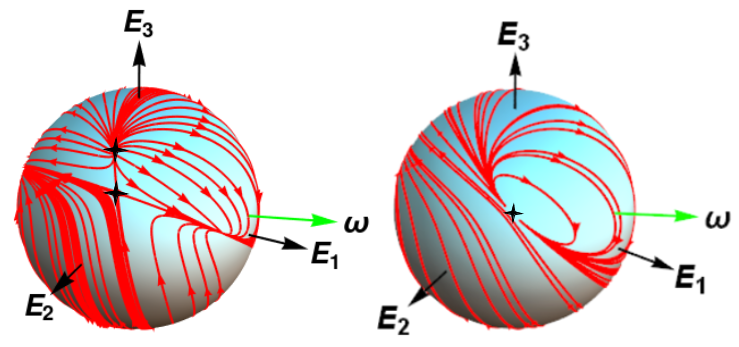}
\vspace*{-0.1in}\caption{Surface streamlines of (a) eccentric hyperbolic\,($\alpha\!\approx\!2.963, \epsilon\!=\!-2,\theta_\omega\!= 4\pi/9,\phi_\omega\!=\!0$) and (b) parabolic\,($\alpha\!=\!2\sqrt{3}, \epsilon\!=\!-2,\theta_\omega\!=\!\tan^{-1}2\sqrt{2},\phi_\omega\!=\!0$) planar linear flows.}
\label{fig:3}
\end{figure}

In the canonical case, the aforementioned neutral cone opens out into a pair of planes($x_1\!=\!\!0;x_2\! =\!\!0$) orthogonal to the flow plane. For sufficiently large $|{\boldsymbol \omega}|$, any line element precesses about the $x_3$-axis, spending equal times in quadrants of positive\,($x_1 x_2 > 0$) and negative\,($x_1x_2 < 0$) rates of stretch, automatically satisfying the net zero rate of stretch required for a closed elliptical streamline. Reducing $|{\boldsymbol \omega}|$ leads to increasingly elongated ellipses that degenerate to straight lines\,(simple shear) at a threshold $|{\boldsymbol \omega}|$\,(for $\hat{\alpha} = 0)$; canonical hyperbolic flows arise for smaller $|{\boldsymbol \omega}|$. An analogous argument implies the existence of eccentric hyperbolic flows(Fig.\ref{fig:3}a), although one needs to reorient ${\boldsymbol \omega}$, while decreasing its magnitude, so the flow remains planar. Simple shear is replaced by a degenerate linear flow with parabolic streamlines\,\footnote{$\mu_i\!=\!0\,\forall\,i$ for both parabolic linear flows and simple shear. While all orientations in the flow-vorticity($x_1\!\!-\!\!x_3$) plane are eigenvectors for simple shear, parabolic flows have a unique in-plane eigenvector, aligned with the axis of the parabolas.} 
at the threshold(Fig.\ref{fig:3}b).

%\begin{figure}[h]
%\includegraphics[height = 0.6\columnwidth, width=0.75\columnwidth]{fig2.png}
%\includegraphics[height = 0.6\columnwidth, width=0.5\columnwidth]{Fig1bnew.png}
%\caption{Eccentric planar elliptic flows result when a fluid element traverses through the neutral cone such that it experiences no net extension or compression as it completes a orbit. Here the red arrow corresponds to vorticity vector and the axes are principal components of extension.}
%\label{fig:2n}
%\end{figure}

\begin{figure*}
\includegraphics[height = 0.85\columnwidth, width=2\columnwidth]{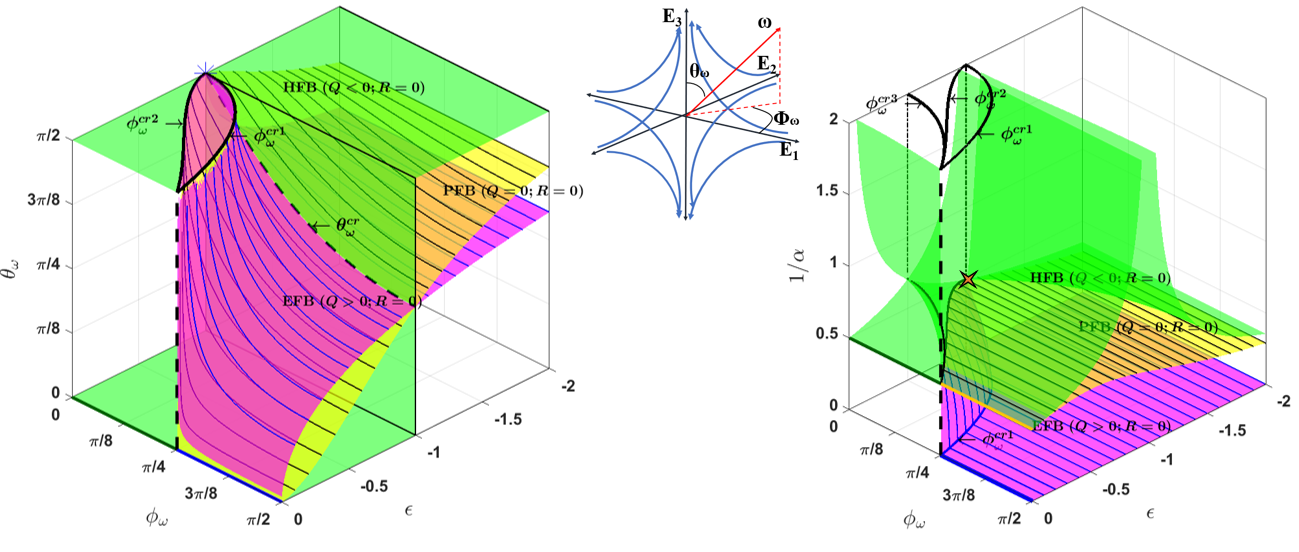}
\vspace*{-0.1in}\caption{The planar linear flow domain in the (a) $\epsilon- \phi_\omega - \theta_\omega$ and (b) $\epsilon- \phi_\omega - 1/\alpha$ sub-spaces. HFB PFB and EFB denote Hyperbolic, Parabolic and Elliptic Flow Boundaries. The central sketch depicts the  flow-type parameters that characterize incompressible linear flows.}\vspace*{-0.25in}
\label{fig:4}
\end{figure*}

Eccentric planar linear flows have $R=0$, with $Q\!>\!0\,(<\!0)$ for eccentric elliptic\,(hyperbolic) flows. So, the $QR$-classification\,\cite{Perry,Cantwell} has canonical and eccentric planar linear flows in identical intervals on the $Q$-axis, highlighting the drawback of a scalar-invariant-based classification - the lack of `structural' information, and the notion of a distance between flows. While this issue may be resolved for canonical compressible planar linear flows\cite{Fairlie,Strogatz} without enlarging the original parameter space\,(see \cite{suppmaterial}), three-dimensional flows require an alternate larger set of structural parameters\cite{Tsinober2009,Meneveau}. A convenient choice for incompressible linear flows is:\!\! $\epsilon\!\!=\!\!E_3/E_2$ measuring the non-axisymmetry of the rate of strain tensor ${\bm E}$;$\alpha\!=\!|\bm{\omega}|/E_2$ measuring the vorticity magnitude (the analog of $2\frac{(1-\hat{\alpha})}{(1+\hat{\alpha})}$ for the canonical case);and angles\,($\theta_\omega,\phi_\omega$)  characterizing the orientation of $\bm{\omega}$ in rate-of-strain-aligned coordinates; see Fig.\ref{fig:4}(center). Incompressible linear flows exist in an $(\epsilon,\alpha,\theta_\omega,\phi_\omega)$-hyperspace with \footnote{One of the eight independent elements of (traceless)\,${\bm \Gamma}$ only acts as an overall scale for the velocity gradient. An additional three degrees of freedom may be eliminated by aligning the coordinate axes with the principal axes of the rate-of-strain tensor. The incompressible linear flow topology is governed by the remaining four degrees of freedom - contrast this with the two scalar parameters in the $QR$-classification}:
\begin{align}
\bm{\Gamma} = \begin{bmatrix}
-(1+\epsilon) & -\frac{\alpha \cos \theta_{\omega}}{2} & -\frac{\alpha \sin \theta_{\omega} \sin \phi_{\omega}}{2}\\
\frac{\alpha \cos \theta_{\omega}}{2} & 1 & -\frac{\alpha \sin \theta_{\omega} \cos \phi_{\omega}}{2}\\
\frac{\alpha \sin \theta_{\omega} \sin \phi_{\omega}}{2} & \frac{\alpha \sin \theta_{\omega} \cos \phi_{\omega}}{2} & \epsilon 
\end{bmatrix}.  \label{Gamma:exp}
\end{align}
Accounting for axes relabeling, and invariance to an overall sign change, all flow topologies are covered for $\epsilon\!\in\![-2,0]$, $\alpha\!\geq\!0$, $\theta_\omega,\phi_\omega \!\in\![0,\!\pi/2]$; $(\epsilon,\alpha,\theta_\omega,\phi_\omega)\!\equiv\!(0,\alpha,0,\phi_\omega)$ and $(-1,\alpha,\frac{\pi}{2},0)$ correspond to canonical planar linear flows, with $\alpha = 2$ for simple shear; $\alpha = 0$ yields linear extensional flows parameterized by $\epsilon$\cite{Deepak1}, $\epsilon\!=\!-2$ being axisymmetric extension. 

We characterize planar linear flows in the above hyperspace, beginning with expressions for the invariants:
\begin{align}
&Q\!=\!\frac{\alpha^2}{4} - (1 + \epsilon + \epsilon^2), \label{Eq.2} \\
&\hspace*{-0.2in}R\!=\!\frac{\epsilon(\!8(1\! +\!\epsilon)\!\! -\!\frac{\alpha^2}{2} (1 \!+\!\!3\cos \theta_{\omega})\!) \!+\! \!(2\!+\!\epsilon)\alpha^2\! \cos 2\phi_{\omega} \!\sin^2\! \theta_{\omega}}{8}\!. %\!+\!  \frac{}{16}
 \label{Eq.3} 
\end{align}
%\begin{figure}
%\includegraphics[height = 0.6\columnwidth,width=0.9\columnwidth]{fig2.png}
%\caption{Eccentric elliptic planar flows result when a fluid element traverses through the neutral cone such that it experiences no net extension or compression as it completes a orbit. Here the red arrow cooresponds to vorticity vector and the axes are principal components of extension.}
%\label{fig:2}
%\end{figure}

Planar linear flows exist in the subspace $R = 0$ corresponding to:
\begin{align}
 \alpha \!&= \!\!\frac{4\sqrt{\epsilon(1+\epsilon)}}{[\cos 2\theta_\omega(3\epsilon \!+\! (2\!+\!\epsilon\!)\!\cos\!2 \phi_\omega)\!\! + \!\!(\epsilon \!-\!(2 \!+\! \epsilon\!)\!\cos \!2\phi_\omega)]^{\frac{1}{2}}}. \label{Eq.7}
\end{align}
The inability to visualize the volume defined by (\ref{Eq.7}), in a 4D-hyperspace, leads us to depict the domain of planar linear flows via its projections onto  $\epsilon\!-\!\phi_\omega\!-\!\theta_\omega$ and $\epsilon\!-\!\phi_\omega\!-\!1/\alpha$ subspaces(Figs.\ref{fig:4}a and b).
%\begin{figure}
%\includegraphics[height = 0.75\columnwidth,width=0.9\columnwidth]{Fig3a_new.png}
%\caption{The $\epsilon- \phi_\omega - \theta_\omega$ sub-space showing the surfaces bounding the domain of existence of eccentric planar linear flows.}
%\label{fig:3}
%\end{figure}
%\begin{figure*}
%\includegraphics[height = 0.925\columnwidth,width=2\columnwidth]{Fig.3_new.png}
%\caption{The domain of planar linear flows in the (a) $\epsilon- \phi_\omega - \theta_\omega$ and (b) $\epsilon- \phi_\omega - 1/\alpha$ sub-spaces.}
%\label{fig:3}
%\end{figure*}

Assuming one boundary of $R =0$ to correspond to solid-body rotation, and using $1/\alpha = 0$ in (\ref{Eq.7}), leads to 
\begin{align}
 \theta^{cr1}_\omega &=\frac{1}{2}\cos^{-1} \left( \frac{(2+\epsilon)\cos 2\phi_\omega - \epsilon}{3\epsilon + (2 + \epsilon)\cos 2\phi_\omega} \right), \label{Eq.8}
\end{align}
which is real-valued for $\phi_\omega \geq \phi_\omega^{cr1}\!(\epsilon)\!=\! \frac{1}{2} \!\cos^{-1}\!\! \left( \!\frac{-\epsilon}{2 + \epsilon} \!\right)$ for $\epsilon\!\!\in\!\![-1,0]$; $\theta^{cr1}_\omega\!=\!\frac{\pi}{2}$ for $\phi_\omega \!=\!\phi_\omega^{cr1}\!(\epsilon)$. The curves [$\theta_\omega\!=\! \frac{\pi}{2},\phi_\omega^{cr1}\!(\epsilon)]$ and $[1/\alpha\! =\!0,\phi_\omega^{cr1}\!(\epsilon)]$, for $\epsilon\!\in\![-1,0]$, therefore mark the edge of the parts of the boundary, comprising solid-body rotation, in the $\epsilon\!-\!\phi_\omega\!-\!\theta_\omega$ and $\epsilon\!-\!\phi_\omega\!-\!1/\alpha$ subspaces. The boundary for $\phi_\omega < \phi_\omega^{cr1}\!(\epsilon)$ consists of finite-$\alpha$ eccentric elliptic flows. It is the plane $\theta_\omega\!=\!\frac{\pi}{2}$ in the former subspace, and given by:
\begin{align}
 \frac{1}{\alpha^{cr1}} = \frac{\sqrt{\epsilon + (2+\epsilon)\cos 2\phi_\omega}}{2\sqrt{-2\epsilon(1+\epsilon)}}, \label{Eq.10}
\end{align}
in the latter one\,(obtained by setting $\theta_\omega\!=\!\frac{\pi}{2}$ in (\ref{Eq.7})). In both subspaces, the boundary ends in an intersection with the parabolic-flow surface obtained next. The curves of intersection are $[\theta_\omega\!=\!\frac{\pi}{2},\phi_\omega^{cr2}\!(\epsilon)]$ and $[1/\alpha\!=\!1/(2\!\sqrt{1\!+\!\epsilon\!+\! \epsilon^2}),\phi_\omega^{cr2}\!(\epsilon)]$, $\phi_\omega^{cr2}(\epsilon)$ being defined below (\ref{Eq.5}).

Figs.\ref{fig:4}a and b show the `elliptic-flow  boundaries' above as magenta surfaces in the $\epsilon\!-\!\phi_\omega\!-\!\theta_\omega$ and $\epsilon\!-\!\phi_\omega\!-\!1/\alpha$  subspaces, respectively, along with the limiting curves\,(in black) $\phi_\omega^{cr1}\!(\epsilon)$ and $\phi_\omega^{cr2}\!(\epsilon)$. For $\epsilon\!\rightarrow \!-2$, the boundary approaches a $\phi_\omega$-independent limit in both subspaces, consistent with axisymmetry of ${\bm E}$. For $\epsilon\rightarrow\!0$,  $\theta^{cr1}_\omega\! \rightarrow\!0$ for $\phi_\omega\in(\frac{\pi}{4},\frac{\pi}{2}]$. For $\phi_\omega = \frac{\pi}{4}$, the limiting $\theta^{cr1}_\omega$ depends on the direction of approach to $[\frac{\pi}{4},0]$ in the $\phi_\omega\!-\!\epsilon$ plane, and ranges over $[0,\frac{\pi}{2}]$ for $\frac{(\phi_\omega - \frac{\pi}{4})}{|\epsilon|}\!\in\![-\frac{1}{4},\infty)$. The elliptic-flow boundary in Fig.\ref{fig:4}a therefore asymptotes to an $L$-shape for $\epsilon\! \rightarrow\!0$\,(dashed black and solid blue lines). Similar arguments show an approach to an $L$-shape in Fig.\ref{fig:4}b, in the same limit, with $1/\alpha \in\![0,\frac{1}{2}]$.

The surface of degenerate parabolic flows, that separates the elliptic and hyperbolic linear flows, is obtained by setting $R = Q = 0$, giving:
\begin{align}
 1/\alpha &= 1/(2\sqrt{1 + \epsilon + \epsilon^2}), \label{Eq.4} \\[5pt]
 \theta_\omega^{cr2} &= \frac{1}{2}\! \cos^{-1}\! \!\left( \!\!1 \!-\! \frac{4 \epsilon^3}{(1 \!+\! \epsilon \!+\! \epsilon^2)(3\epsilon\! +\! (2\! +\! \epsilon)\!\cos 2\phi_\omega)}\! \right)\!, \label{Eq.5}
\end{align}
which correspond to the yellow surfaces in Fig.\ref{fig:4}. $\theta_\omega^{cr2}$ is real-valued for
$\phi_\omega \!\!\geq\! \phi_\omega^{cr2}\!(\epsilon)\!=\! \frac{1}{2}\!\cos^{-1}\!\!\left(\!\frac{3\epsilon + 3\epsilon^2 - \epsilon^3}{2 + 3\epsilon + 3\epsilon^2 + \epsilon^3} \!\!\right)$ for $\epsilon\!\in\!(-1,0)$.  $\theta^{cr2}_\omega\!\!\rightarrow\!0$, for $\epsilon\!\!\rightarrow\!0$, for $\phi_\omega\!\in\!\!(\frac{\pi}{4},\frac{\pi}{2}]$, but ranges over $\!\![0,\frac{\pi}{2}]$ depending on the ratio $\frac{(\phi_\omega - \frac{\pi}{4})/|\epsilon|+\frac{3}{4}}{\epsilon^2}$. The parabolic surface in Fig.\ref{fig:4}a therefore asymptotes to the same $L$-shape, as the elliptic flow boundary, for $\epsilon\!\rightarrow\!0$. The parabolic and elliptic-flow boundaries, in Fig.\ref{fig:4}a, also  intersect along the (dashed black)\,curve $\theta_\omega^{cr1}\!\!=\!\!\theta_\omega^{cr2}\!\!=\!\!\theta_\omega^{cr}\!=\!\! \frac{1}{2}\!\cos^{-1}\! \!\left(\!\frac{\cos 2\phi_\omega\!+\!1}{\cos 2\phi_\omega\!-\!3}\! \right)$  in the plane $\epsilon\!=\!-1$.

The hyperbolic-flow boundary, consisting almost entirely of finite-$\alpha$ flows, requires analyzing the zero-crossings of $R$ v/s $\alpha$ curves, details of which are given in \cite{suppmaterial}. For a given $(\epsilon,\phi_\omega)$, a hyperbolic flow on the boundary corresponds to the zero-crossing at the smallest $\alpha$\,(with $Q\!<\!0$). For $\phi_\omega\!\!>\!\! \phi_\omega^{cr1}\!(\epsilon)$, there is a single zero-crossing at a finite $\alpha$, for $\theta_\omega\!\in\!(\theta_\omega^{cr1},\frac{\pi}{2})$ when $\epsilon\!\!\in\! [-2,-1)$, and for $\theta_\omega\!\in\!(0,\theta_\omega^{cr1})$ when $\epsilon\!\in\!(-1,0]$. $R$ is independent of $\alpha$ at $\theta_\omega\!=\!\theta_\omega^{cr1}$, corresponding to a zero-crossing at infinity\,(solid-body rotation). The $\alpha$ at the zero crossing decreases\,(increases) with increasing $\theta_\omega$ for $-2\!<\!\epsilon\!<\!1$\,($-1\!<\! \epsilon\!<\!0$), so the hyperbolic-flow boundary is $\theta_\omega\!=\!\frac{\pi}{2}$ for $\epsilon\!<\!- 1$ and $\theta_\omega\!=\!0$ for $\epsilon\!>\!-1$. This boundary, shown in green in Fig.\ref{fig:4}a, lies above the elliptic-flow boundary for $\epsilon\!<\!-1$, but below it for $\epsilon\!>\!-1$. Using these $\theta_\omega$'s  in (\ref{Eq.7}) leads to the two pieces of the hyperbolic-flow boundary in Fig.\ref{fig:4}b for $\phi_\omega \geq \phi_\omega^{cr1}\!(\epsilon)$, given by (\ref{Eq.10}) for $\epsilon\!\!\in\!\! [-2,-1)$ and by $1/\alpha^{cr2}\!=\!1\!/\!(2\sqrt{1\!+\!\epsilon})$ for $\epsilon\!\!\in\!\! (-1,0]$. Both bounding surfaces\,(green) diverge for $\epsilon\!\rightarrow\!-1$, planar extension($\alpha=0$) being the limiting flow in this plane.

$\theta_\omega^{cr1}$ not being real-valued for $\phi_\omega < \phi_\omega^{cr1}\!(\epsilon)$, with $\epsilon\!\in\!(-1,0)$, implies the absence of a zero-crossing at $\alpha = \infty$. As mentioned above, this leads to a lifting of the elliptic-flow boundary, across $\phi_\omega\!=\!\phi_\omega^{cr1}(\epsilon)$, onto the finite-$\alpha$ surface, defined by (\ref{Eq.10})\,(Fig.\ref{fig:4}b); the hyperbolic-flow boundary remains $\theta_\omega = 0$ for $\phi_\omega^{cr2}\!(\epsilon) \!\leq\! \phi_\omega \!< \!\phi_\omega^{cr1}\!(\epsilon)$. For $\phi_\omega < \phi_\omega^{cr2}(\epsilon)$, all zero-crossings of $R$ have $Q\!<\!0$, and all planar linear flows are hyperbolic. Thus, in the region of the $\epsilon-\phi_\omega$ plane corresponding to $\epsilon\!\!\in\!\!(-1,0)$,$\phi_\omega \!<\! \phi_\omega^{cr2}\!(\epsilon)$, the domain of planar linear flows is contained between hyperbolic-flow boundaries given by $\theta_\omega\!=\!\frac{\pi}{2}$ and $0$. $\theta_\omega\!=\!\frac{\pi}{2}$ is a continuation of the elliptic-flow boundary, defined by (\ref{Eq.10}), to $\phi_\omega\!<\!\phi_\omega^{cr2}(\epsilon)$, and is the third piece of the hyperbolic-flow boundary in Fig.\ref{fig:4}b being connected to the piece in $\epsilon\!\!\in\!\![-2,-1)$ at  $(\epsilon,\phi_\omega,1/\alpha)\!\equiv\!(-1, 0, \frac{1}{2})$. Interestingly, the two hyperbolic-flow boundaries above intersect along $\phi_\omega^{cr3}(\epsilon)\!=\!\cos^{-1}(-\frac{3\epsilon}{2+\epsilon})$ in Fig.\ref{fig:4}b, so planar hyperbolic flows along this curve correspond to a unique $\alpha\,(= 2\sqrt{1\!+\!\epsilon})$; see \cite{suppmaterial}.

Sets of vanishing measure in the subspaces above correspond to canonical planar linear flows. The point $(\epsilon,\phi_\omega,\theta_\omega)\!\equiv\!(-1,0,\frac{\pi}{2})$ in Fig.\ref{fig:4}a includes all canonical flows, with $\alpha$ determining the flow-type. The horizontal arm of the limiting $L$-form of the elliptic and parabolic-flow surfaces, given by $[\epsilon\!=\!0,\theta_\omega\!=\!0,\phi_\omega\!\in\![\frac{\pi}{4},\frac{\pi}{2}]]$, contains the canonical elliptic flows and simple shear; it maps onto the $\epsilon\!=\!0$ plane with $\phi_\omega\!\in\![\frac{\pi}{4},\frac{\pi}{2}]$,$1/\alpha\!\in\![0,\frac{1}{2}]$ in Fig.\ref{fig:4}b. Canonical hyperbolic flows and simple shear occupy the complementary interval $[\epsilon\!=\!0,\theta_\omega\!=\!0,\phi_\omega\!\in\![0,\frac{\pi}{4}]]$ in Fig.\ref{fig:4}a\,(dark green line) and the part of the $\epsilon = 0$ plane with $\phi_\omega\!\in\![0,\frac{\pi}{4}]$,$1/\alpha\!\in\![\frac{1}{2},\infty)$ in Fig.\ref{fig:4}b. Points on the vertical arm of the $L$, defined by $[\epsilon\!=\!0, \phi_\omega \!=\!\frac{\pi}{4},\theta_\omega\!\in\!(0,\frac{\pi}{2})]$, and on the critical curve $[\epsilon\!\!=\!\!-1, \theta_\omega^{cr}\!(\phi_\omega)]$, include all three eccentric planar-flow-types. Any path approaching $\theta_\omega\!\neq\!\theta^{cr}_\omega(\phi_\omega)$ on $\epsilon\!=\!-1$, or $\theta_\omega\!\neq\!0$ on $\epsilon\!=\!0$\,(with $\phi_\omega < \frac{\pi}{4}$), asymptotes to planar extension\,($1/\alpha\!\rightarrow\!\infty$). In this limiting sense, the planes $\epsilon\!=\!-1$ and $\epsilon\!=\!0\,(\phi_\omega\!<\!\frac{\pi}{4})$, outside of the above curves, may be regarded as parts of the hyperbolic-flow boundary \footnote{Both the disconnected nature of the  bounding surfaces in Fig.\ref{fig:4}, and the associated degeneracies, are an artifact of the lower-dimensional projections.}. %In the $\epsilon- \phi_\omega - 1/\alpha$  sub-space (Fig.\ref{fig:3}(b)), the surface (yellow surface corresponding to Eq.\ref{Eq.4}) asymptotes to the line $1/\alpha = 1/2$ as $\epsilon \rightarrow 0$.
%\begin{figure}
%\includegraphics[height = 0.85\columnwidth,width=0.9\columnwidth]{Fig3b_new.png}
%\caption{The $\epsilon- \phi_\omega - 1/\alpha$ sub-space showing the surfaces bounding the domain of existence of eccentric planar linear flows.}
%\label{fig:4}
%\end{figure}
%\begin{figure*}
%\includegraphics[height = 0.85\columnwidth,width=1\columnwidth]{Fig4.png}
%\includegraphics[height = 0.85\columnwidth,width=1\columnwidth]{Fig5.png}
%\caption{The $\epsilon- \phi_\omega - \theta_\omega$ sub-space showing the special planes (a) $\epsilon = -1$ and (b) $\epsilon = 0$, highlighting the planar flows that populate these planes.}
%\label{fig:4}
%\end{figure*}
\vspace{0.1in}

Returning to Fig.\ref{fig:2}, the surface streamlines in the canonical case(Fig.\ref{fig:2}b) are Jeffery orbits, the curves of intersection of the unit sphere with a one-parameter family of right elliptical cones given by $x_1^2\!-\!\hat{\alpha} x_2^2 \!\!=\!\! C^2\!x_3^2$\cite{Mason}. This leads to the well known orbit constant $C\!\!=\! \frac{\tan \theta (\kappa^2\! \cos^2\!\phi\! +\! \sin^2\!\phi)^{1/2}}{\kappa}$ that parameterizes the orbits; $C\!=\!0$ and $\infty$ are the spinning and tumbling orbits\cite{Hinch1,Hinch2}. Here, $\kappa\!\!=\!\! 1/\!\sqrt{-\hat{\alpha}}$, and $\theta$ and $\phi$ are angles in a spherical coordinate system with its polar axis aligned with the axis of the cone family. Interestingly, the surface streamlines in the eccentric case\,(Fig.\ref{fig:2}a) are generalized Jeffery orbits, obtained as curves of intersection of the unit sphere with the family of inclined elliptical cones,  $(x_1\!-\!\hat{\delta}x_3)^2\!+\!\frac{(x_2\!-\! \hat{\gamma}x_3)^2}{(1-e^2)}\!=\!\tilde{C}^2\!x_3^2$, where
%\begin{widetext}
\vspace*{-0.125in}
\begin{multline}
\hspace*{-0.2in}\begin{split}
 \tilde{C}\!=\!\tan \theta \Big( \Big.\frac{1}{1\!-\!e^2} \Big[\!\Big.(\!\sin^2 \phi\!-\!\hat{\gamma}^2)\!\!+\!\csc^2 \theta (\hat{\gamma}^2\!-\!\hat{\gamma}\!\sin \phi \sin 2\theta) \Big. \Big]  \\
\hspace*{-0.2in}+\!\!\left[\!(\!\cos^2 \phi\!-\!\hat{\delta}^2)\!\!+\!\csc^2 \theta (\hat{\delta}^2\!\!-\!\hat{\delta}\!\cos \phi \sin 2\theta) \right] \Big.\!\!\Big)^{\!\!\frac{1}{2}}, \label{Eq.11}
 \end{split} 
\end{multline}
%\vspace*{-0.22in}
%\end{widetext}
is the generalized orbit constant. The polar axis defining $(\theta,\phi)$ is chosen normal to the flow plane; $\hat{\gamma}$, $\hat{\delta}$ and $e$ in (\ref{Eq.11}) are expressible in terms of flow-type parameters; $\hat{\gamma} = \hat{\delta} = 0$, $e\!=\!(1\!+\! \hat{\alpha})^{1/2}$ gives the canonical case\,(see \cite{suppmaterial}). 

The significance of the eccentric planar flows is first seen for scalar transport from a drop in an ambient linear flow, a problem ubiquitous in nature and industry; the transport rate  at large Peclet numbers($Pe$) is governed by the near-surface streamline topology. Surface streamlines are projections of the streamlines of an auxiliary linear flow whose ${\bm \Gamma}$ differs from (\ref{Gamma:exp}) in diagonal elements having an additional factor $(1\!+\!\lambda)^{-1}$, $\lambda$ being the drop-to-medium viscosity ratio\!\cite{Deepak1}. The auxiliary flow generically projects onto an open surface-streamline topology, with the dimensionless transport rate\,(the Nusselt number $Nu$) scaling as $Pe^{\frac{1}{2}}$ for $Pe\!\rightarrow\!\infty$\cite{Deepak1,Deepak2}. For canonical planar flows, however, the auxiliary flow has closed streamlines for all $\lambda$ when $\hat{\alpha}\!<\!0$, and for $\lambda\!>\!\frac{2\hat{\alpha}}{(1\!-\!\hat{\alpha})}$ when $\hat{\alpha}\!>\!0$\cite{Deepak1,Powell}. The resulting closed surface-streamline topology implies diffusion-limited transport-\!$Nu$ approaches an $O(1)$ value for $Pe\!\rightarrow\!\infty$\cite{Deepak1,Deepak2}. A closed surface-streamline topology must, in fact, arise whenever the auxiliary flow is an eccentric elliptic flow. Crucially, unlike the canonical case, the near-surface streamlines are open but tightly spiralling, similar to those around a rigid particle in a vortical linear flow\cite{Batchelor1}. Thus, drops in linear flows, with associated auxiliary flows that are eccentric elliptic flows, mimic rigid particles with $Nu\!\propto\!Pe^{\frac{1}{3}}$\cite{Batchelor1,Goddard, Sub1,Sub2}! The asymptotic 'transport surface' obtained by plotting $Nu/Pe^{\frac{1}{2}}$ as a function of flow-type parameters and $\lambda$\cite{Deepak1,Deepak2}, exhibits a singular dip along the locus of auxiliary eccentric elliptic flows\cite{Sabarish}. The actual and auxiliary  flows are identical for bubbles\,($\lambda\!=\!0$) which must exhibit diffusion limitation\,($Nu\!\sim\!O(1)$) in canonical elliptic flows, but convective enhancement\,($Nu\!\propto\!Pe^{\frac{1}{3}}$) in eccentric elliptic flows.

A second scenario concerns the evolution of an orientable microstructure in a linear flow, as governed by $\dot{\boldsymbol p}\!\!=\!\bm{\Omega}\!\cdot\!\bm{p}\! +\!B (\bm{E}\!\cdot\!\bm{p}\!-\!(\bm{E\!:\!pp})\bm{p})$; $\bm{\Omega}$ being the vorticity tensor and ${\bm{p}}$ the microstructure orientation. For spheroids of aspect ratio $\kappa$, $B\!=\!\frac{\kappa^2 - 1}{\kappa^2 + 1}$, and the curve $\hat{\alpha}\!\!=\!\!\kappa^{-2}(\kappa^2)$ demarcates flow-aligning and tumbling dynamics of prolate\,(oblate) spheroids in the $\kappa\!-\!\hat{\alpha}$ plane for canonical planar flows\cite{Hinch1,Marath2}; tumbling occurs along Jeffery orbits\cite{Hinch1}. The analogous problem for eccentric planar flows\cite{suppmaterial} shows flow-aligning behavior on either side of the critical curve for finite $\kappa$, with the approach to alignment changing character\,(spiralling vs non-spiralling); generalized Jeffery orbits only arise for $\kappa\!=\!0$ and $\!\infty$. One may extend these considerations to polymer molecules - it is of interest to examine the nature of the coil-stretch transition\cite{deGennes,hinch77,ChuShaqfeh2003,Shaqfeh2,HoffmanShaqfeh2007} along eccentric planar linear flow sequences.

Planar linear flows have been examined in the context of hydrodynamic instabilities \cite{PanThien,Kerswell} and turbulence \cite{HuntConf,Hunt}. Canonical planar flows, both hyperbolic and elliptic \cite{Kerswell,Ledizes1999}, serve as approximate base-states for the short-wavelength instabilities of stretched vortices with a precise alignment between the axes of extension and vorticity. The eccentric planar flows enable one to circumvent this alignment approximation, potentially allowing insight into the short-wavelength dynamics of a more general class of vortical structures.

%\begin{figure}\phi_
%\includegraphics[scale = 0.21]{Fig_4_new.png}
%\caption {A representative plot of the critical curve separating the spiralling and non-spiralling trajectories of prolate spheroids (generalised Jeffery orbits) for $\epsilon = -2$, $\omega = 0$ and $\alpha = \alpha^{cr}$. The trajectories are shown in the inset for a finite $\kappa = 10$ (left) and $\kappa = \infty$ (right).}
%\label{fig:4}
%\end{figure}

%The expression (Eq.\ref{Eq.11}) above corresponds to the special case of the trajectories traced by infinite aspect ratio spheroids in eccentric elliptic flows, and one can extend it to the case of arbitrary aspect ratio $\kappa$.
%The original $QR$-classification has been successful in answering important questions such as the bias of sub-Kolmogorov (linear)\,flows, even in homogeneous isotropic turbulence, towards a biaxial extensional topology - a sampling of these flows leads to a characteristic tear-drop-shaped region in the $QR$-plane\,\cite{Meneveau}. The sub-Kolmogorov flow topology has important implications for the stretching dynamics of polymeric molecules, drop break-up and coalescence, etc. Nevertheless, as already mentioned above, this classification does not differentiate between the canonical and eccentric planar linear flows. As will be seen below, there are important problems for which the answers depend on the overall structure of the flow, and are therefore different for the canonical and eccentric cases, in turn  requiring the alternate $(\epsilon,\alpha,\theta_\omega,\phi_\omega)$-based classification above.

\begin{acknowledgments}
SVN acknowledges the financial support of JNCASR, Bangalore, India.
\end{acknowledgments}
%%%%%%%%%%%%%%%%%%%%%%%%%%%%%%%%%%%%%%%%%%%%%%%%%%%%%%%%%%%%%%%

\end{document}

\section{Supplemental Material}
%\maketitle  
\section{Canonical planar linear flows}
Canonical planar linear flows are a one-parameter subset of the four-parameter family of incompressible linear flows. Defining the linear flow field as ${\boldsymbol u} = {\boldsymbol \Gamma} \cdot {\boldsymbol x}$, and writing ${\boldsymbol \Gamma} = {\boldsymbol E} + {\boldsymbol \Omega}$, one has ${\boldsymbol E} = \frac{(1+\hat{\alpha})}{2}\!\left[\!\begin{smallmatrix} 0 & 1 & 0 &\\ 1 & 0 & 0 \\ 0 & 0 & 0\!\end{smallmatrix}\!\right]$ and ${\boldsymbol \Omega} = \frac{(1-\hat{\alpha})}{2}\!\left[\!\begin{smallmatrix} 0 & 1 & 0 &\\ -1 & 0 & 0 \\ 0 & 0 & 0\!\end{smallmatrix}\!\right]$ for the rate of strain and vorticity tensors associated with the canonical planar linear flows; the vorticity vector is normal to the $x_1-x_2$ plane, being given by ${\boldsymbol \omega}= - \bm{\epsilon} \mathbf{:} \bm{\Omega} = (\hat{\alpha}-1){\boldsymbol 1}_3$. The parameter $\hat{\alpha}$ may therefore be written in terms of the characteristic magnitudes of extension\,($E = \frac{(1+\hat{\alpha})}{2}$) and rotation\,($\Omega = \frac{(1-\hat{\alpha})}{2}$) as $\hat{\alpha}= \frac{E - \Omega}{E + \Omega}$; alternatively, one has $E/\Omega = \frac{(1+\hat{\alpha})}{(1-\hat{\alpha})}$ for the ratio of (in-plane)\,extension to (out-of-plane)\,vorticity. In Fig.\ref{fig:s0o} below, we plot both the 3D streamline patterns\,(blue and black) and their projections\,(red-dashed) onto the unit sphere, the surface streamlines, for all three classes of canonical flows. The streamlines for canonical elliptic linear flows with $-1 \leq \hat{\alpha} < 0$,  shown on the left, are ellipses; those for canonical hyperbolic linear flows with $0 < \hat{\alpha} \leq 1$, shown on the right, are hyperbolae. The ellipses degenerate to circles for solid-body rotation\,($\hat{\alpha} = -1$), and the hyperbolae reduce to rectangular hyperbolae for planar extension\,($\hat{\alpha} = 1$). The surface streamlines for the canonical elliptic and hyperbolic flows are non-planar versions of the corresponding plane quadratic curves; those for elliptic flows in particular turn out to have the same form as Jeffery orbits which are a one-parameter family of closed curves, on the unit sphere, traced by the (unit)\,orientation vector of a neutrally buoyant freely rotating axisymmetric particle in simple shear flow\cite{Jeffery}. Simple shear flow with $\hat{\alpha} = 0$, shown in the center of Fig.\ref{fig:s0o}, has straight\,(rectilinear) streamlines which project onto a meridional surface-streamline configuration with the pair of intersections of the unit sphere with the flow direction serving as the polar points. Figure \ref{fig:s0o} also shows the organization of the canonical flows along the $Q$-axis\,($R=0$) as per the $QR$-classification\cite{Perry,Cantwell}.
\begin{figure*}[h]
\centering
\includegraphics[scale=0.45]{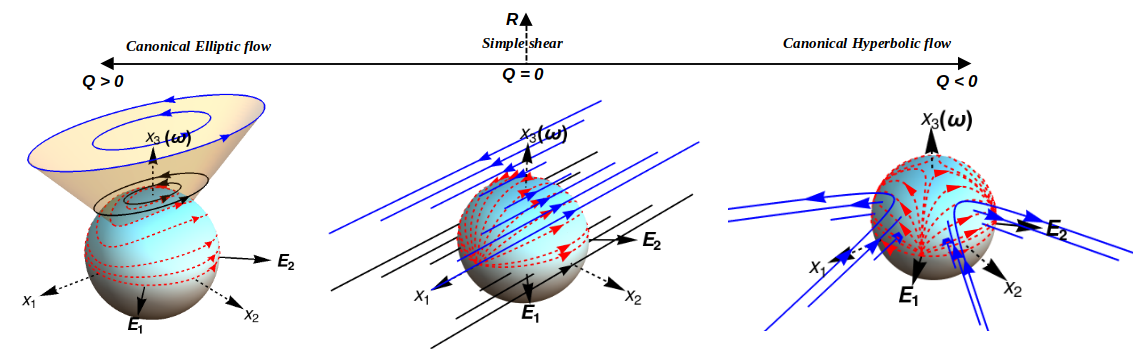}
\vspace*{-0.15in}\caption{The $Q$-axis with the intervals corresponding to the different canonical planar flows\,(above). The surface\,(red-dashed) and 3D\,(blue/black) streamline patterns for the canonical elliptic planar linear flows\,(left), the hyperbolic planar linear flows\,(right) and simple shear flow\,(center); $\bm{\omega}$ is the vorticity vector, and $E_{1,2}$ are the principal in-plane extensions\,($E_3=0$).}\vspace*{-0.3in}
\label{fig:s0o}
\end{figure*}

\section{Classification of 2D Compressible linear flows}

In this section, we present an improved two-parameter classification scheme for compressible planar linear flows. The original scalar-invariant-based classification for such flows, the so-called $PQ$-scheme, $P$ here being the negative trace of ${\boldsymbol \Gamma}$,  lacks the notion of a distance between the different planar linear flow topologies\cite{Perry}; this scheme is the projection, onto $R = 0$, of the more general three-parameter $PQR$-scheme for 3D linear flows\cite{Cantwell}. The $PQ$-scheme succeeds in demarcating flow topologies with spiralling and non-spiralling streamlines across a critical curve\,(a parabola defined by $P^2 = 4Q$, and obtained from equating the quadratic discriminant to zero), and in addition, locates the incompressible canonical planar linear flows along the $Q$-axis\,($P=0$) with simple shear flow at the origin\,($Q=P =0$); the $P$-axis\,($Q=0$) corresponds to degenerate saddle-node topologies\,(owing to one of the eigenvalues being zero). However, the scheme does not specify precise locations either for the canonical hyperbolic and elliptic flows along the $Q$-axis, or for aforementioned degenerate topologies along the $P$-axis\,($Q = 0$), or for the compressible flows within the different regions demarcated by $Q = 0$ and $P^2 = 4Q$. 

\begin{figure}[H]
\centering
\includegraphics[scale = 0.7]{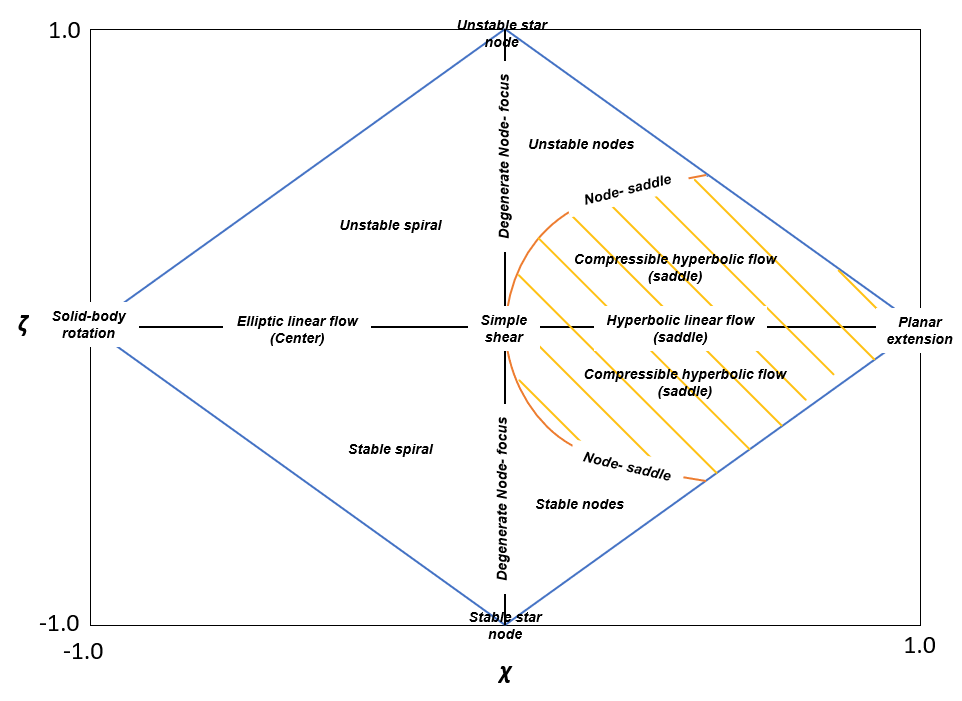}
\caption{The $\chi - \zeta$ classification scheme for 2D planar linear flow topologies.}
\label{fig:s3}
\end{figure}

Instead of basing one's classification on scalar invariants, one may begin instead from the following physical decomposition of the (transpose of the)\,velocity gradient tensor:
\begin{align}
    \bm{\Gamma} = \zeta 
    \begin{bmatrix}
    1 &0 \\
    0 &1
    \end{bmatrix} + (1 - |\zeta|)
    \left( \chi 
    \begin{bmatrix}
    1 &0 \\
    0 &-1
    \end{bmatrix} + (1 - \chi)
    \begin{bmatrix}
    0 &1 \\
    -1 &0
    \end{bmatrix}
    \right). \label{Eq.8}
\end{align}
The three $2$\,X\,$2$ matrices in (\ref{Eq.8}) correspond to pure dilatation, planar extension and solid-body rotation. Thus, the parameter $\zeta$ is a normalized measure of dilatation or compressibility, with the parameter $\chi$, for a fixed $\zeta$, playing a role similar to $\hat{\alpha}$ for the incompressible canonical linear flows, in terms of determining the relative magnitudes of rate of (inplane)\,strain and (out-of-plane)\,vorticity\,($\chi = \frac{(1+\hat{\alpha})}{2}$). One may organize the planar linear flows in the $\chi-\zeta$ plane, based on (\ref{Eq.8}), the domain being a square with boundaries $\chi = \pm 1$, $\zeta = \pm 1$. However, such a classification leads to a degeneracy: each of the pair of boundaries $\zeta = \pm 1$ correspond to the same linear flow topology - pure dilatation\,(a star-node configuration). In order to remove this degeneracy, we re-write (\ref{Eq.8}) in the form:
\begin{widetext}
\begin{align}
    \bm{\Gamma} = \zeta 
    \begin{bmatrix}
    1 &0 \\
    0 &1
    \end{bmatrix} + (1 - |\zeta|)
    \left( \frac{(2 \chi - |\zeta| + 1}{2(1 - |\zeta|)} 
    \begin{bmatrix}
    1 &0 \\
    0 &-1
    \end{bmatrix} + (1 - \frac{(2 \chi - |\zeta| + 1}{2(1 - |\zeta|)})
    \begin{bmatrix}
    0 &1 \\
    -1 &0
    \end{bmatrix}
    \right) \label{Eq.9}
\end{align}
\end{widetext}
where $\chi$ in (\ref{Eq.8}) is now replaced by $\frac{(2 \chi - |\zeta| + 1}{2(1 - |\zeta|)}$. The transformation amounts to making the length of the $\zeta$-interval depend on $\chi$ in such a manner that the upper and lower boundaries, $\zeta = \pm 1$, associated with (\ref{Eq.8}), now shrink to the points $(\chi,\zeta) = (0,\pm 1)$; further, (\ref{Eq.9}) also ensures that the simple shear flow topology corresponds to the origin, $(\chi,\zeta) \equiv (0,0)$, as was the case with the $P-Q$ scheme. One may readily identify the different planar linear flow topologies based on a straightforward analysis of the eigenvalues and eigenvectors associated with the matrix in (\ref{Eq.9}), and the detailed organization of these flow topologies is provided in Fig.\ref{fig:s3}. The critical curve separating the spiralling and non-spiralling streamline topologies, the analog of $P^2 = 4Q$ in the original scheme, is now the vertical segment $\chi = 0,|\zeta| \leq 1$. The horizontal segment $\zeta = 0,|\chi| \leq 1$ is populated by incompressible canonical planar linear flows, and is the analog of the $Q$-axis in the original scheme. Finally, the orange curve in Fig.\ref{fig:s3}, given by $\zeta^2 + 2 \xi(|\zeta| - 1) = 0$, is populated by degenerate saddle-node topologies and demarcates the region populated by nodes outside from that populated by saddles within; this curve is the analog of the $P$-axis in the original classification.

\section{The hyperbolic-flow boundary in the $\epsilon - \phi_{\omega} - \theta_\omega$ and $\epsilon - \phi_{\omega} - 1/\alpha$ subspaces}
\begin{figure*}[h]
\vspace{-0.1in}
\includegraphics[scale = 0.465]{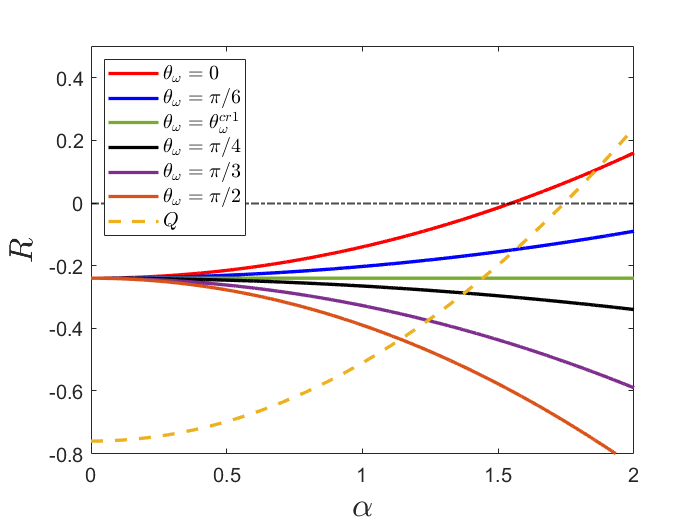}
\includegraphics[scale = 0.465]{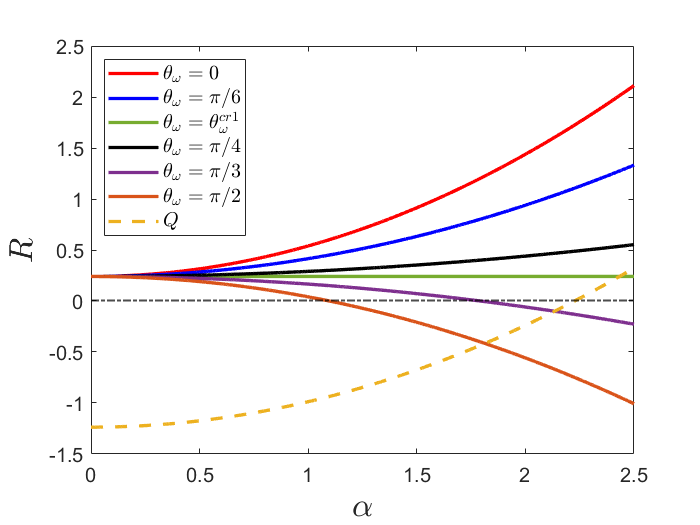}
\vspace{-0.1in}
\caption{Plot of the cubic invariant $R$ as a function of $\alpha$ for different $\theta_{\omega}$, for the \textbf{Type 1} scenario: (a) The curves are for $(\epsilon,\phi_\omega) \equiv (-0.4,\frac{\pi}{3})$, $\phi_\omega^{cr1} = \frac{1}{2}\cos^{-1}(\frac{1}{4}) \approx 37.8^\circ$, and are representative of $\epsilon$'s in the interval $-1 < \epsilon < 0$; (b) the curves are for $(\epsilon,\phi_\omega) \equiv (-1.1,\frac{\pi}{3})$, $\phi_\omega^{cr1} = 0$, and are representative of $\epsilon$'s in the interval $-2 < \epsilon < 1$. The dashed curves in (a) and (b) correspond to the quadratic invariant $Q$, given by equation (2) in the main manuscript, and is independent of both $\phi_\omega$ and $\theta_\omega$.}
\label{fig:s0}
\end{figure*}

\begin{figure}[H]
\centering
\includegraphics[scale = 0.45]{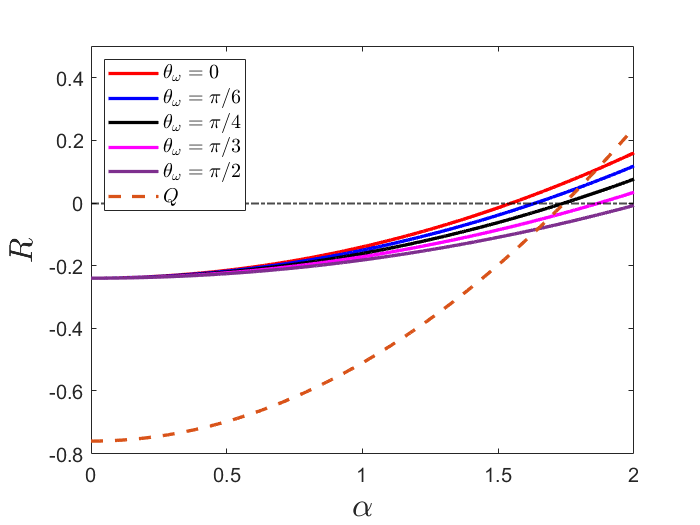}
\includegraphics[scale = 0.45]{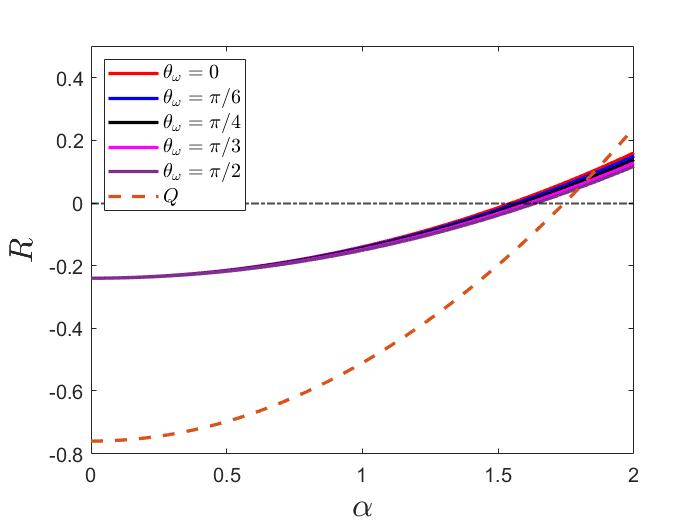}\\
\includegraphics[scale = 0.45]{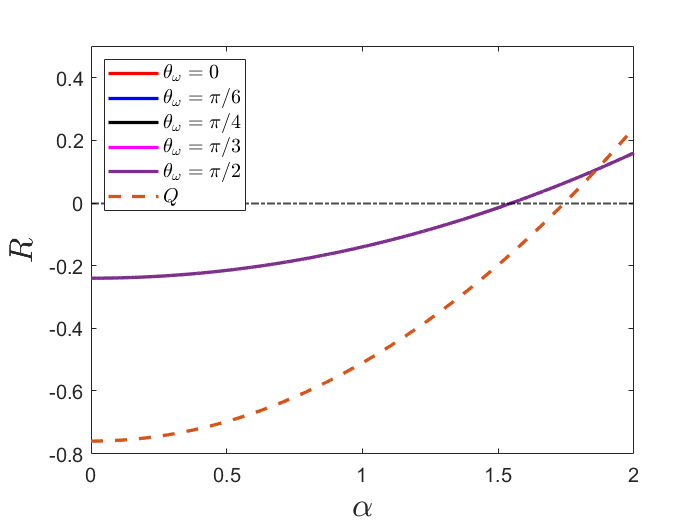}
\includegraphics[scale = 0.45]{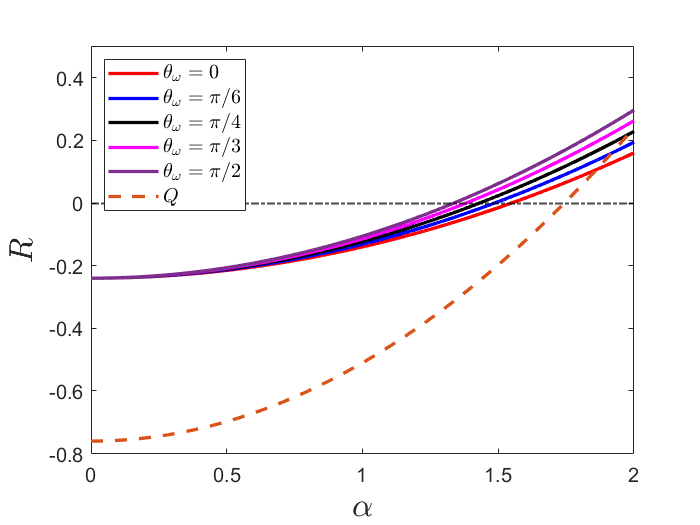}
\caption{Plots of the cubic invariant $R$ as a function of $\alpha$ for different $\theta_{\omega}$ for the \textbf{Type 2} scenario ($\epsilon = -0.4$, $\phi_{\omega} < \phi_{\omega}^{cr1} \approx 38^{\circ}; \phi_{\omega}^{cr2} \approx 25^{\circ}; \phi_{\omega}^{cr3} \approx 20^{\circ}$): (a) $\phi_{\omega} = 28^{\circ}$ ($\phi_{\omega}^{cr1} > \phi_{\omega} > \phi_{\omega}^{cr2}$); (b) $\phi_{\omega} = 22^{\circ}$ ($\phi_{\omega}^{cr2} > \phi_{\omega} > \phi_{\omega}^{cr3}$); (c) $\phi_{\omega} = \phi_{\omega}^{cr3} \approx 20^{\circ}$; (d) $ \phi_{\omega} = 15^{\circ} (\phi_{\omega} < \phi_{\omega}^{cr3})$. The dashed curve in all cases corresponds to the quadratic invariant $Q$.}
\label{fig:s1}
\end{figure}

Herein, we describe, in more detail, the construction of the hyperbolic-flow boundary in the $\epsilon-\phi_{\omega} - \theta_\omega$ and $\epsilon - \phi_{\omega} - 1/\alpha$ subspaces - that is, the piecewise smooth green surfaces shown in Fig.3 of the main manuscript. The construction of this bounding surface for the latter subspace is more involved since it requires a detailed analysis of the dependence of the cubic invariant $R$ on $\alpha$. Since $R = 0$ for planar linear flows, such flows correspond to zero crossings of the $R$\,\,v/s\,\,$\alpha$ curves, with the limiting hyperbolic planar linear flow, for a given $(\epsilon,\phi_\omega)$, corresponding to the zero crossing at the smallest $\alpha$ with $Q < 0$. The nature of the $R$\,\,v/s\,\,$\alpha$ curves is a function of the flow-type parameters, and there emerge different regimes for the behavior of the zero crossings for $\phi_\omega > \phi_{\omega}^{cr1}(\epsilon)$ and  $\phi_\omega < \phi_{\omega}^{cr1}(\epsilon)$; these are shown in Figs.\ref{fig:s0} and \ref{fig:s1}, respectively. Prior to the detailed analysis below, two points are worth mentioning: (1) All of the eccentric hyperbolic planar linear flows have a non-zero vorticity; the only irrotational planar linear flow is therefore planar extension; (2) Eccentric hyperbolic planar linear flows can, however, have a planar rate-of-strain field similar to the canonical flows, the only difference being that the vorticity vector is no longer orthogonal to the plane of strain.  

Figs.\ref{fig:s0}a and b plot $R$ as a function of $\alpha$ for different $\theta_{\omega}$'s, for a given $\phi_\omega > \phi_\omega^{cr1}(\epsilon)$, and for a pair of $\epsilon$'s in the intervals $-1 < \epsilon < 0$ and $-2 < \epsilon < -1$, respectively; the expression used for $R$ is equation (3) in the main manuscript. In  Fig.\ref{fig:s0}a, $R$ transitions from a monotonically increasing to a monotonically decreasing function of $\alpha$, as $\theta_{\omega}$ increases from $0$ to $\frac{\pi}{2}$, with the value of $R$ at any given $\alpha$ decreasing monotonically with increasing $\theta_\omega$. Since $R$ starts off at a negative value for $\alpha = 0$, a zero-crossing exists as long as $R$ is a monotonically increasing function of $\alpha$. This zero-crossing occurs at the smallest $\alpha$ for $\theta_\omega = 0$, moving to progressively larger $\alpha$ with increasing $\theta_\omega$, eventually diverging for $\theta_\omega \rightarrow \theta_\omega^{cr1-}$; for $\theta_\omega = \theta_\omega^{cr1}$, defined in equation (5) of the manuscript, $R$ is independent of $\alpha$ and appears as a horizontal line. The above behavior occurs for any $\epsilon$ in the interval $(-1,0)$, and implies that the limiting hyperbolic planar flow corresponds to $\theta_\omega = 0$ for $\epsilon$'s in this interval. The other limiting flow, the one on the elliptic-flow boundary, is a solid-body rotation for $\theta_\omega \rightarrow \theta_\omega^{cr1-}$. The identification of flow-type\,(hyperbolic vis-a-vis elliptic) is consistent with the associated sign of $Q$ - this is clarified by the plot of the quadratic invariant $Q$ in Figs.\ref{fig:s0}a and b. The behavior of the  $R$\,\,v/s\,\,$\alpha$ curves in Fig \ref{fig:s0}b is similar to that described above, except that $R$ now starts off at a positive value for $\alpha = 0$. As a result, the zero-crossing is now restricted to cases where $R$ is a monotonically decreasing function of $\alpha$; it occurs at the smallest $\alpha$ for $\theta_\omega = \frac{\pi}{2}$, moving to larger $\alpha$ with decreasing $\theta_\omega$, eventually diverging for $\theta_\omega \rightarrow \theta_\omega^{cr1+}$. The limiting hyperbolic flow now corresponds to $\theta_\omega = \frac{\pi}{2}$, with the limiting flow on the elliptic-flow boundary, for $\theta_\omega \rightarrow \theta_\omega^{cr1+}$, again being a solid body rotation. Note that there are no planar linear flows for $\theta_\omega > \theta_\omega^{cr1}$ when $-1 < \epsilon < 0$, and for  $\theta_\omega < \theta_\omega^{cr1}$ when $-2 < \epsilon < 1$.

For the scenario above corresponding to $\phi_\omega > \phi_\omega^{cr1}(\epsilon)$, which we label the \textbf{Type 1} scenario, the hyperbolic-flow boundary corresponds to $\theta_\omega = 0$ for $\epsilon\,\in\,(-1,0)$ and to $\theta_\omega = \frac{\pi}{2}$ for $\epsilon\,\in\,(-2,-1)$ - these two planes are shown in green in Fig.3a of the main manuscript. The elliptic flow boundary always corresponds to solid-body rotation in this range, lying below the hyperbolic flow boundary for $\epsilon\,\in\,(-2,-1)$, and above it for $\epsilon\,\in\,(-1,0)$. The zero-crossing of the $R$-curve, corresponding to the smallest $\alpha$, always lies to the left of that of the $Q$-curve, so that this zero crossing, along with the one at infinity\,(solid-body rotation), brackets the point $Q = 0$. Hence, in the above scenario, for a given $(\epsilon,\phi_\omega)$, one traverses the hyperbolic planar linear flows, and then the elliptic planar linear flows\,(separated by an intervening degenerate parabolic flow at $\theta_\omega = \theta_\omega^{cr2}$, with $\theta_\omega^{cr2}$ being defined by equation (8) in the main manuscript), with increasing $\theta_\omega$ for $\epsilon\,\in\,(-1,0)$; and with decreasing $\theta_\omega$ for $\epsilon\,\in\,(-2,-1)$.

One may obtain the hyperbolic-flow boundary in the $\epsilon - \phi_\omega - 1/\alpha$ subspace, for the \textbf{Type 1} scenario, by substituting the value of $\theta_\omega$ corresponding to the smallest-$\alpha$ zero crossing in the expression (4), in the main manuscript, that defines the subspace $R = 0$. As mentioned above, this is $\theta_{\omega} = \pi/2$ for $\epsilon\,\in\,(-2,-1)$, and $\theta_{\omega} = 0$ for $\epsilon\,\in\,(-1,0)$, $\phi_\omega > \phi_\omega^{cr1}(\epsilon)$, and one obtains:
    \begin{equation}
    \frac{1}{\alpha}=\begin{cases}
          \frac{\sqrt{\epsilon + (2 + \epsilon) \cos 2\phi_{\omega}}}{2 \sqrt{2}\sqrt{-\epsilon(1+\epsilon)} } \quad &\text{for}\,\, \epsilon\,\in\,(-2,-1), \\
          \frac{1}{2\sqrt{1 + \epsilon}} \quad &\text{for}\,\,\epsilon\,\in\,(-1,0),\,\phi_\omega > \phi_\omega^{cr1}(\epsilon),  \\
     \end{cases} \label{Type1:alpha}
    \end{equation}
which correspond to the two green surfaces in Fig.3b of the main manuscript in the region $\phi_\omega > \phi_\omega^{cr1}(\epsilon)$. Note that the elliptic-flow boundary in the $\epsilon - \phi_\omega - 1/\alpha$ subspace, for $\phi_\omega > \phi_\omega^{cr1}(\epsilon)$, is just $1/\alpha = 0$, corresponding to solid-body rotation.

The second \textbf{Type 2} scenario, corresponding to $\phi_\omega < \phi_\omega^{cr1}(\epsilon)$\,(implying $\epsilon\,\in\,(-1,-0)$), is characterized by the plots in  Fig.\ref{fig:s1}. Unlike the \textbf{Type 1} scenario above, there does not exist a threshold $\theta_\omega$ corresponding to an $\alpha$-independent $R$, and all four sets of plots in Fig.\ref{fig:s1}a-d show $R$ to be a monotonically increasing function of $\alpha\,\,\forall\,\, \theta_\omega\,\in\,[0,\frac{\pi}{2}]$. The absence of a zero-crossing at infinity, corresponding to an $\alpha$-independent $R$, implies that the limiting flow on the elliptic-flow boundary is no longer a solid-body rotation. Therefore, for a given $(\epsilon,\phi_\omega)$, the planar linear flows on both bounding surfaces are finite-$\alpha$ flows\,(that is, have a finite ratio of vorticity to extension). Further, the plots in Fig.\ref{fig:s1} show that the $R$\,\,v/s\,\,$\alpha$ curves, and thence, the associated interval of zero-crossings, move across the $Q$\,\,v/s\,\,$\alpha$ curve, towards smaller $\alpha$, with decreasing $\phi_\omega$. Interestingly, this is accompanied by a reversal in the order of the zero-crossings. The reversal occurs across a critical $\phi_\omega = \phi_\omega^{cr3}(\epsilon)$, at which the $R$-curves for the different $\theta_\omega$ collapse onto a single one, and the expression for which may therefore be obtained by setting $dR/d\theta_\omega = 0$, which gives $\phi_{\omega}^{cr3}(\epsilon) = \frac{1}{2} \cos^{-1}\left( \frac{-3 \epsilon}{2 + \epsilon}\right)$. The smallest-$\alpha$ zero-crossing corresponds to $\theta_\omega = 0$ for $\phi_\omega > \phi_{\omega}^{cr3}(\epsilon)$, and to $\theta_\omega = \frac{\pi}{2}$ for $\phi_\omega < \phi_{\omega}^{cr3}(\epsilon)$. 

For $\phi_{\omega}^{cr1}(\epsilon) > \phi_{\omega} > \phi_{\omega}^{cr2}(\epsilon)$, as exemplified by the curves in Fig.\ref{fig:s1}a, the point $Q = 0$ lies within the interval of $R$\,\,v/s\,\,$\alpha$ zero crossings. This is analogous to the Type 1 scenario above\,(for $\epsilon\,\in,(-1,0))$ with there being hyperbolic planar linear flows followed by elliptic planar linear flows, for $\theta_\omega$ increasing from $0$ to $\frac{\pi}{2}$. The only difference is the absence of a zero crossing at infinity. So, while the limiting hyperbolic flow, corresponding to the smallest-$\alpha$ zero crossing, still has $\theta_\omega = 0$, the other limiting flow corresponding to the largest-$\alpha$ zero crossing is now a finite-$\alpha$ eccentric elliptic flow with $\theta_\omega = \frac{\pi}{2}$. The hyperbolic-flow boundary remains $\theta_\omega = 0$ in the $\epsilon -\phi_\omega - \theta_\omega$ subspace, and is still given by $1/\alpha = 1/(2\sqrt{1+\epsilon})$ in the $\epsilon -\phi_\omega - 1/\alpha$ subspace\,(the second of the expressions in (\ref{Type1:alpha}) above). The difference arises for the elliptic-flow boundary in the two subspaces: in the $\epsilon -\phi_\omega - \theta_\omega$ subspace, this boundary coincides with the plane $\theta_\omega = \frac{\pi}{2}$, while in the $\epsilon -\phi_\omega - 1/\alpha$ subspace, it lifts off from the plane corresponding to solid-body rotation\,($1/\alpha = 0$) along $\phi_\omega = \phi_{\omega}^{cr1}(\epsilon)$, the subsequent form of this lifted boundary being obtained by using $\theta_\omega = \frac{\pi}{2}$ in expression (4) of the main manuscript, leading to the first expression in (\ref{Type1:alpha}) above.

\begin{figure}[H]
\includegraphics[scale = 0.355]{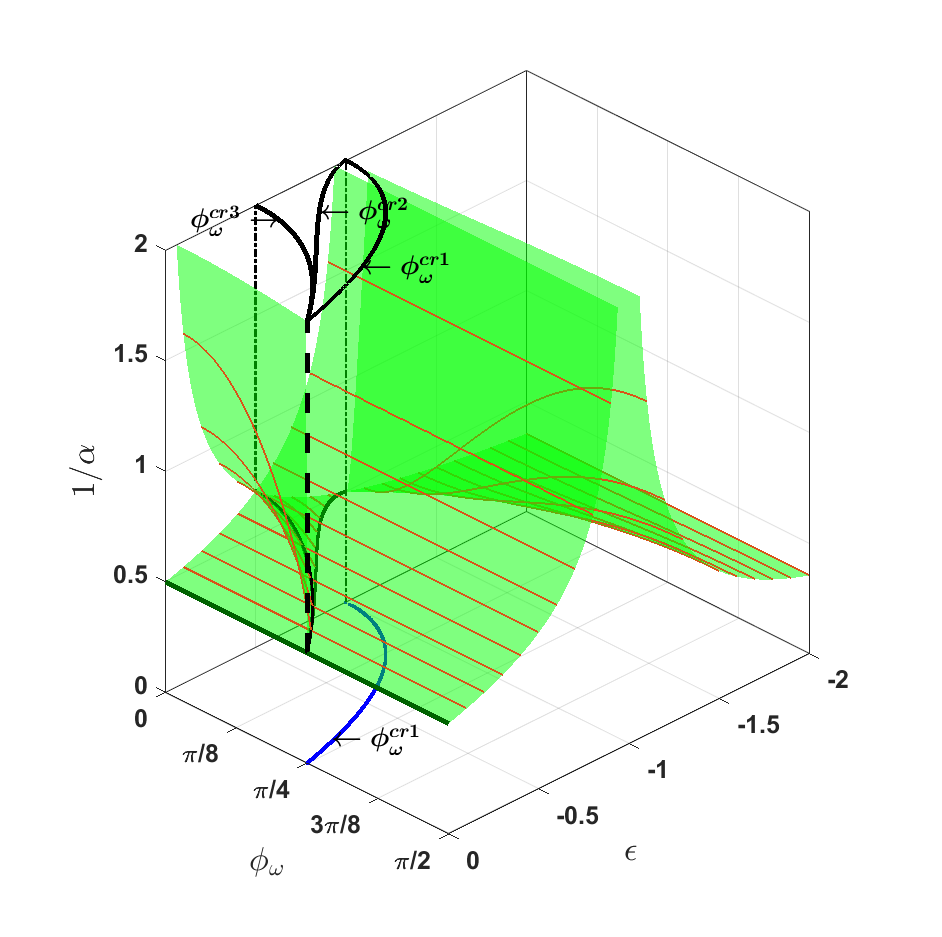}
\includegraphics[scale = 0.355]{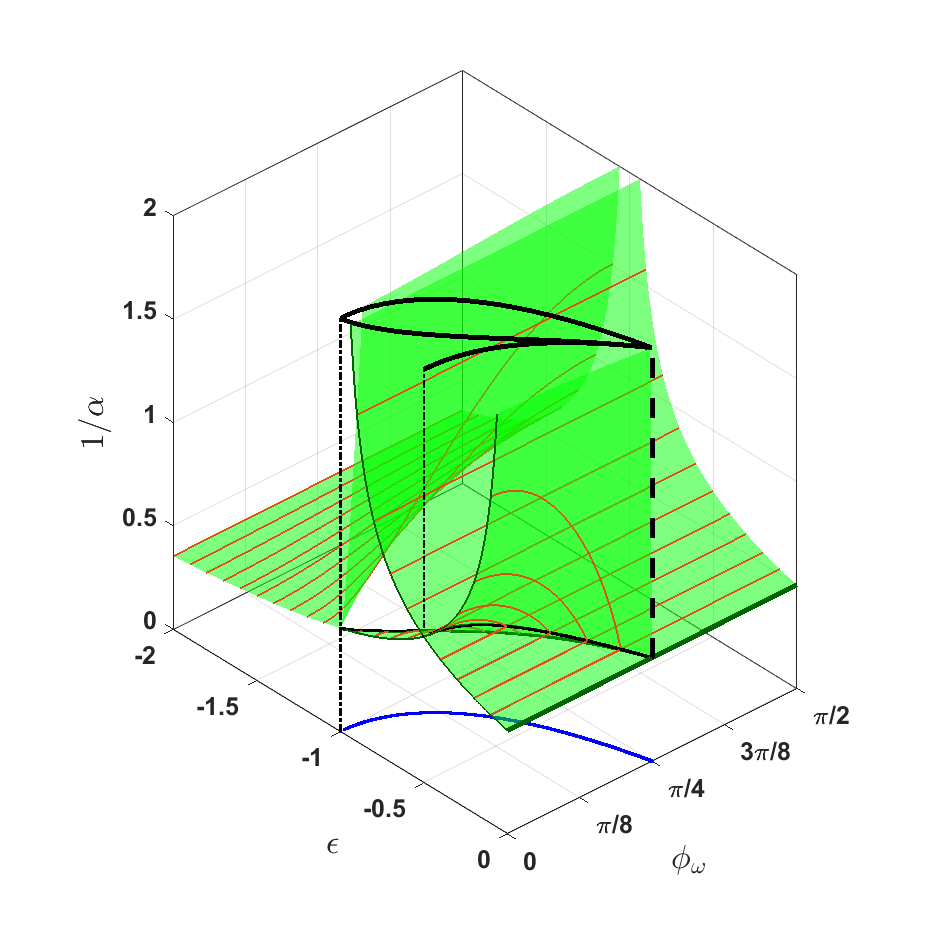}\\
\centering
\includegraphics[scale = 0.45]{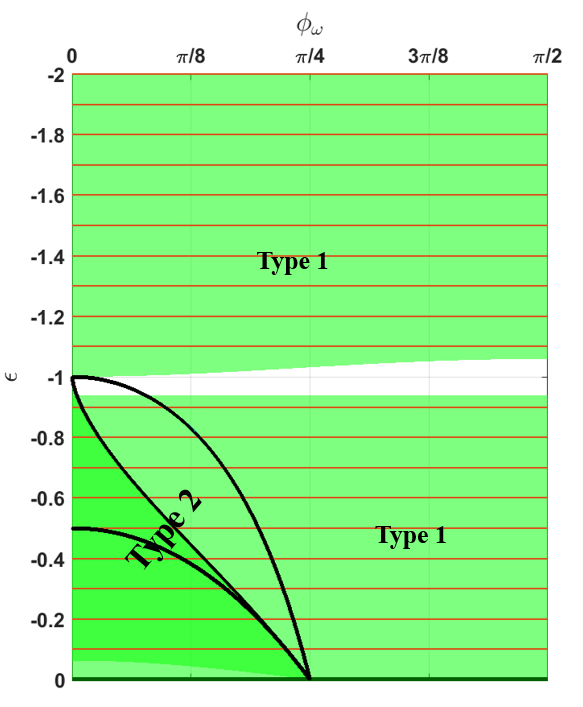}
\caption{The hyperbolic flow boundary in the $\epsilon-\phi_\omega-1/\alpha$ subspace.}
\label{fig:s1a}
\end{figure}

For the interval $\phi_\omega^{cr2}(\epsilon) > \phi_\omega > \phi_\omega^{cr3}(\epsilon)$ characterized by the curves in Fig.\ref{fig:s1}b, all of the $R$\,\,v/s\,\,$\alpha$ zero crossings have moved to the left of $Q = 0$, and the limiting planar flows corresponding to both the smallest and largest zero-crossings are now hyperbolic flows with $Q < 0$; these zero crossings still correspond to $\theta_\omega = 0$ and $\frac{\pi}{2}$, respectively. Thus, for a given $(\epsilon,\phi_\omega)$, the interval of planar linear flows is contained between the $\theta_\omega = 0$ and $\theta_\omega = \frac{\pi}{2}$ planes, which now correspond to the upper and lower hyperbolic-flow boundaries in the $\epsilon - \phi_\omega - \theta_\omega$ subspace, with the latter being the continuation of the elliptic-flow boundary in the interval $\phi_{\omega}^{cr1}(\epsilon) > \phi_{\omega} > \phi_{\omega}^{cr2}(\epsilon)$ above; these are shown as the upper and lower green planes in Fig.3a of the main manuscript for $\phi_\omega < \phi_\omega^{cr2}(\epsilon)$. For $\phi_\omega = \phi_\omega^{cr3}(\epsilon)$, the $R\,\,v/s\,\,\alpha$ curves for the different $\theta_\omega$'s collapse onto a single curve, and the associated zero-crossing corresponds to $Q< 0$, as shown in Fig.\ref{fig:s1}c. For smaller $\phi_\omega$ pertaining to the interval $0 < \phi_\omega < \phi_\omega^{cr3}(\epsilon)$, the smallest and largest zero-crossings of the $R$\,\,v/s\,\,$\alpha$ curves correspond to $\theta_\omega = \frac{\pi}{2}$ and $0$, respectively. While the hyperbolic flow boundaries remain unaltered in the $\epsilon - \phi_\omega - \theta_\omega$ subspace, the existence of the aforementioned critical value, $\phi_{\omega}^{cr3}(\epsilon)$, implies an intersection of the upper and lower hyperbolic-flow boundaries in the $\epsilon -\phi_\omega - 1/\alpha$ subspace along the (projected) curve $\phi_\omega = \phi_\omega^{cr3}(\epsilon)$, accompanied an exchange of identities. Thus, one has:
   \begin{equation}
    \frac{1}{\alpha_{min}}=\begin{cases}
          \frac{\sqrt{\epsilon + (2 + \epsilon) \cos 2\phi_{\omega}}}{2 \sqrt{2}\sqrt{-\epsilon(1+\epsilon)} } \quad &\text{if} \, \phi_{\omega} < \phi_{\omega}^{cr3}(\epsilon), \\
          \frac{1}{2\sqrt{1 + \epsilon}} \quad &\text{if} \, \phi_\omega^{cr2}(\epsilon) > \phi_{\omega} > \phi_{\omega}^{cr3}(\epsilon), \\
     \end{cases} \label{Eq.4}
    \end{equation}
    \begin{equation}
    \frac{1}{\alpha_{max}}=\begin{cases}
          \frac{\sqrt{\epsilon + (2 + \epsilon) \cos 2\phi_{\omega}}}{2 \sqrt{2}\sqrt{-\epsilon(1+\epsilon)} } \quad &\text{if} \, \phi_\omega^{cr2}(\epsilon) > \phi_{\omega} > \phi_{\omega}^{cr3}(\epsilon), \\
          \frac{1}{2\sqrt{1 + \epsilon}} \quad &\text{if} \, \phi_{\omega} < \phi_{\omega}^{cr3}(\epsilon). \\
     \end{cases} \label{Eq.5}
    \end{equation}
As a consequence, and remarkably, along the curve $\phi_\omega = \phi_\omega^{cr3}(\epsilon)$, the entire interval of hyperbolic planar linear flows has the same $1/\alpha$ despite $\theta_\omega$ varying from $0$ to $\frac{\pi}{2}$; this common value corresponds to the latter of the two expressions in (\ref{Type1:alpha}), as must be the case since this expression is independent of $\phi_\omega$.  
Different views of the hyperbolic flow boundary pertaining to the $\epsilon-\phi_\omega-1/\alpha$ subspace, and obtained based on the above arguments, are shown in Fig.\ref{fig:s1a}. It is of interest to clarify the limiting form of this boundary for $\epsilon\rightarrow\!0$, corresponding to a planar rate-of-strain field. Similar to the elliptic and parabolic surfaces in the  $\epsilon-\phi_\omega-\theta_\omega$ subspace, described in the manuscript, the limiting form of the hyperbolic flow boundary in the $\epsilon-\phi_\omega-1/\alpha$ subspace  depends on $\phi_\omega$. All three threshold curves, $\phi_\omega^{cr1}(\epsilon)$, $\phi_\omega^{cr2}(\epsilon)$ and $\phi_\omega^{cr3}(\epsilon)$, approach $\frac{\pi}{4}$ for $\epsilon \rightarrow 0$, and the limiting form of the hyperbolic boundary therefore differs in the intervals $\phi_\omega \in [0,\frac{\pi}{4})$ and $\phi_\omega \in (\frac{\pi}{4},\frac{\pi}{2}]$, and right at $\phi_\omega = \frac{\pi}{4}$. The piece of the boundary given by $1/\alpha = 1/2\sqrt{1+\epsilon}$ - the second expression in (\ref{Type1:alpha}) and (\ref{Eq.5}) - trivially approaches $1/2$ in the limit $\epsilon \rightarrow 0$ for $\phi_\omega\in[0,\frac{\pi}{2}]$; this corresponds to simple shear flow. This is the only piece of the hyperbolic flow boundary for $\phi_\omega \in (\frac{\pi}{4}, \frac{\pi}{2}]$, and bounds the domain of planar linear flows from above; it merges with the parabolic surface in this interval. The aforesaid piece is the lower boundary for $\phi_\omega \in [0,\frac{\pi}{4})$, with the upper boundary in this $\phi_\omega$-interval being given by the first expression for $\frac{1}{\alpha_{min}}$ in (\ref{Eq.4}), and that diverges to infinity for  $\epsilon \rightarrow 0$. For $\phi_\omega = \frac{\pi}{4}$, the limiting form of this upper boundary depends on the direction of approach to $[\frac{\pi}{4},0]$ in the $\epsilon\!-\!\phi_\omega$ plane, and ranges over $[\frac{1}{2},\infty)$ for $\frac{(\phi_\omega - \frac{\pi}{4})}{|\epsilon|}\!\in\![-\frac{1}{4},\infty)$. As a result, the two pieces of the hyperbolic flow boundary asymptote to a `$\perp$' sign for $\epsilon \rightarrow 0$, with the horizontal arm corresponding to $1/\alpha = 1/2$ in this plane. The region above this arm\,($1/\alpha > 1/2$), for $\phi_\omega \in [0,\frac{\pi}{4})$ corresponds to canonical hyperbolic flows, while the region below the arm\,($1/\alpha < 1/2$), for $\phi_\omega \in (\frac{\pi}{4},\frac{\pi}{2}]$ corresponds to canonical elliptic flows, with the horizontal arm itself corresponding to simple shear, as mentioned above. Interestingly, the vertical arm of the `$\perp$' are eccentric hyperbolic flows with the vorticity vector not being orthogonal to the plane of strain.
 %Fig. 6 thereafter depicts this boundary along with the elliptic and parabolic bounding surfaces; the different perspective complements the one shown in Fig.3 of the main manuscript.

\section{Generalised Jeffery Orbits - Surface streamlines of eccentric elliptic linear flows}
It was argued in the manuscript that the surface streamlines of an eccentric elliptic planar linear flow are generalized Jeffery orbits. The associated expression for the generalized orbit constant is given in equation (9) of the manuscript, and was derived using a geometric approach\cite{thesis} in a spherical coordinate system with its polar axis normal to the plane of the flow. The constants $\hat{\gamma}$ and $\hat{\delta}$, and $e$, that appear in this expression, characterize the inclination\,(with the polar axis) of the real eigenvector $\bm{v}_3$ of $\bm{\Gamma}$, and the eccentricity of the elliptical streamlines, respectively. As mentioned in the manuscript, $\bm{v}_3$ corresponds to the zero eigenvalue, and is aligned with the loci of the elliptical streamline centers. One can derive expressions for the aforementioned constants in terms of the flow-type parameters $(\epsilon, \alpha, \theta_\omega, \phi_\omega)$, starting from the equation governing the eigenvalues\,($\mu$) of $\bm{\Gamma}$:
\begin{align}
    \mu^3 + Q \mu + R = 0.
\end{align}
For planar linear flows $R = 0$, and the above equation is readily solved to obtain the three eigenvalues $\mu_{1,2} = \pm \imath \sqrt{Q}$ and $\mu_3 = 0$, with $Q$ given by equation (2) in the manuscript. Using these eigenvalues, one may calculate the eigenvectors as solutions of $(\bm{\Gamma} - \mu_i \bm{I}) \cdot \bm{v}_i = 0$. The neutral eigenvector is given by:
\begin{align}
    \bm{v}_3 = \begin{pmatrix}
               \frac{2 \epsilon  \sin \left(\phi _{\omega }\right) \left(\sin \left(\theta _{\omega }\right)+\epsilon  \cos \left(\theta _{\omega }\right) \cot \left(\theta_{\omega }\right) \csc ^2\left(\phi _{\omega }\right)\right)}{\sqrt{\epsilon  (\epsilon +1)} \sqrt{3 \epsilon  \cos \left(2 \theta _{\omega }\right)-2(\epsilon +2) \sin ^2\left(\theta _{\omega }\right) \cos \left(2 \phi _{\omega }\right)+\epsilon }+2 \epsilon  (\epsilon +1) \cos \left(\theta _{\omega}\right) \cot \left(\phi _{\omega }\right)} \\
              \frac{\epsilon  \cot \left(\theta _{\omega }\right) \csc \left(\phi _{\omega }\right) \sqrt{3 \epsilon  \cos \left(2 \theta _{\omega }\right)-2 (\epsilon +2)\sin ^2\left(\theta _{\omega }\right) \cos \left(2 \phi _{\omega }\right)+\epsilon }-2 \sqrt{\epsilon  (\epsilon +1)} \sin \left(\theta _{\omega }\right)\cos \left(\phi _{\omega }\right)}{\sqrt{3 \epsilon  \cos \left(2 \theta _{\omega }\right)-2 (\epsilon +2) \sin ^2\left(\theta _{\omega }\right) \cos \left(2 \phi _{\omega }\right)+\epsilon }+2 \sqrt{\epsilon  (\epsilon +1)} \cos \left(\theta _{\omega }\right) \cot \left(\phi _{\omega }\right)} \\
              1
              \end{pmatrix}
\end{align}
%\vspace{-0.25in}
where the constraint $R = 0$ has been imposed. One may now find $\hat{\delta}$ and $\hat{\gamma}$ as:
\begin{align}
    \hat{\delta} &= \frac{|(v_3^1)|}{\sqrt{(v_3^1)^2 + (v_3^2)^2 + (v_3^3)^2}} \label{Eq.delta}\\
    \hat{\gamma} &= \frac{|(v_3^2)|}{\sqrt{(v_3^1)^2 + (v_3^2)^2 + (v_3^3)^2}} \label{Eq.gamma}
\end{align}
%\vspace{-0.25in}
where $v_3^i$ denotes the $i^{th}$component of $\bm{v}_3$. Note that the components of $\bm{v}_3$ above are defined in a rate-of-strain aligned coordinate system. Thus, the expressions (\ref{Eq.delta} and (\ref{Eq.gamma}) define the inclination with respect to the third axis of extension, corresponding to the component $E_3$; see expression for $\bm{\Gamma}$ as given in equation (2) of the manuscript.

Similarly, the eccentricity $e$ is given by:
\begin{align}
    e  = \left(1 - \frac{1}{|\mathfrak{Im}(\bm{v}^2)|} \right)^{1/2} \label{Eq.e}
\end{align}
where $|\mathfrak{Im}(\bm{v}^2)|$ is the norm of the imaginary part of the eigenvector $\bm{v}_2$ given by:
\begin{align}
    \bm{v}^2 = \begin{pmatrix}
                1\\
                \frac{-\alpha  \sin (\theta ) \sin (\phi )+\tan (\theta ) \left(i \sqrt{\alpha ^2-4 \left(\epsilon ^2+\epsilon +1\right)}+2 \epsilon +2\right) \cos (\phi)}{\alpha  \sin (\theta ) \cos (\phi )+i \tan (\theta ) \left(\sqrt{\alpha ^2-4 \left(\epsilon ^2+\epsilon +1\right)}+2 i\right) \sin (\phi )}\\
                \frac{\sec (\theta ) \left(-\alpha ^2 \sin ^2(\theta )+2 i \epsilon  \sqrt{\alpha ^2-4 \left(\epsilon ^2+\epsilon +1\right)}+4 \epsilon ^2\right)}{\alpha\left(\alpha  \sin (\theta ) \cos (\phi )+i \tan (\theta ) \left(\sqrt{\alpha ^2-4 \left(\epsilon ^2+\epsilon +1\right)}+2 i\right) \sin (\phi )\right)}
               \end{pmatrix}
\end{align} 
with $\alpha = \frac{2 \sqrt{\epsilon  (\epsilon +1)}}{\sqrt{\epsilon  \cos ^2(\theta )-\frac{1}{2} \sin ^2(\theta ) ((\epsilon +2) \cos (2 \phi )+\epsilon )}}$ from the constraint $R=0$.

\section{Dynamics of an orientable microstructure: Canonical and Eccentric planar linear flow sequences}
\begin{figure}[H]
\vspace{-0.1in}
\includegraphics[scale = 0.35]{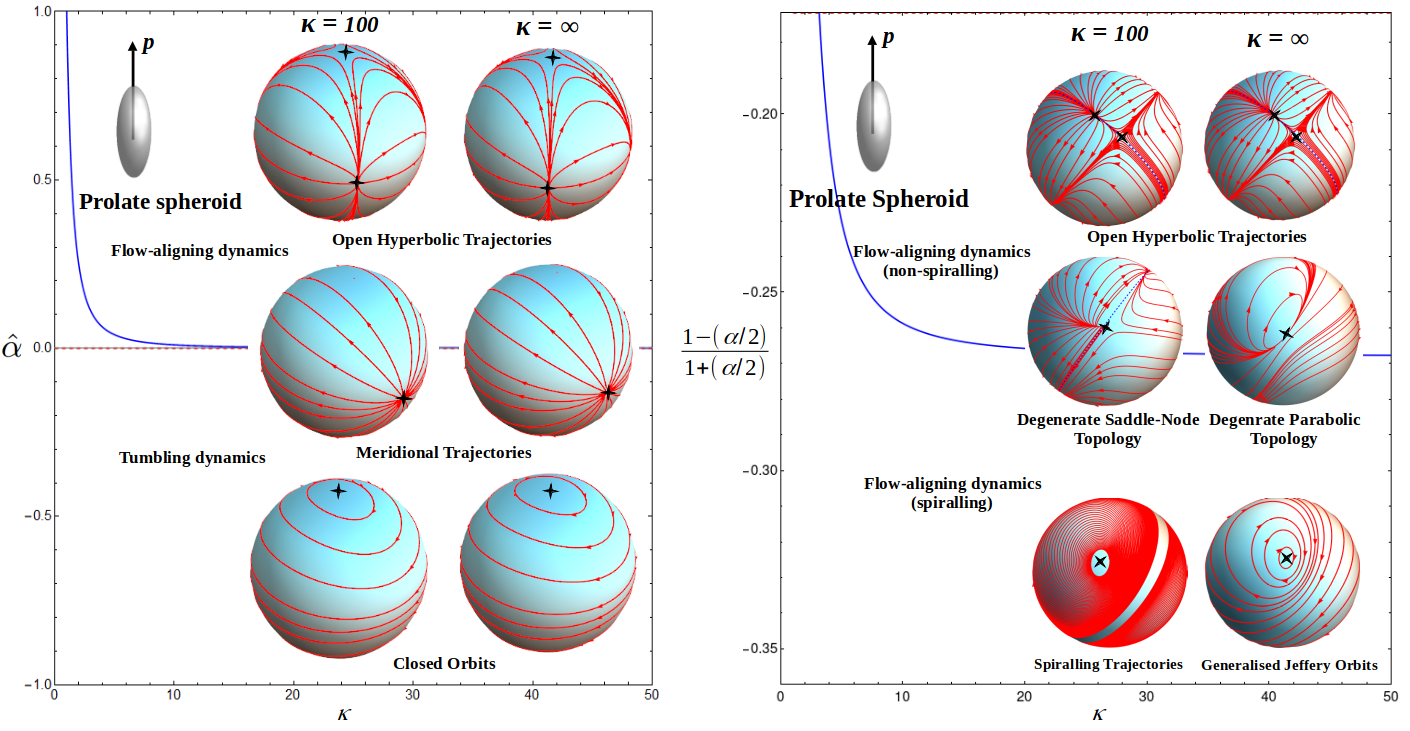}
%\includegraphics[scale = 0.65]{S4b.PNG}
%\vspace{-0.5in}
\caption{Unit-sphere trajectory topologies, for prolate spheroids with $\kappa = 100$ and $\kappa =\infty$, along (a) a canonical linear flow sequence with $\hat{\alpha}$ varying from $-1$ to $1$, and (b) an eccentric linear flows sequence with $\alpha$ varying from $-1$ to $\approx -0.18$. The blue curve in each figure corresponds to $\Delta = 0$, where $\Delta$ is the discriminant of the (transpose of) velocity gradient tensor, $\hat{\bm{\Gamma}} = \bm{\Omega} + \frac{\kappa^2 - 1}{\kappa^2 + 1}\bm{E}$, which enters the auxiliary equation, $\dot{\bm{x}} = \bm{\Gamma \cdot x}$, associated with (\ref{spher:orient}); for the canonical flows, this threshold curve reduces to $\hat{\alpha} = 1/\kappa^2$.}
\label{fig:s4}
\end{figure}

In Figs.\ref{fig:s4}a and b, we plot the trajectories traced on the unit sphere, by the unit orientation vector(${\bm p}$) associated with a prolate spheroid of a varying aspect ratio, for different members of canonical and eccentric planar linear flow sequences. The figures help organize the types of orientation dynamics, as characterized by the unit-sphere trajectory topology, on a plane comprising the spheroid aspect ratio $\kappa$ and an appropriate flow-type parameter; $\kappa > 1$ for prolate spheroids. The prolate spheroid here is taken to be representative of a generic orientable microstructure, with the evolution of its orientation satisfying:
\begin{align}
    \bm{\dot{p}} = \bm{\Omega \cdot p} + \frac{\kappa^2 - 1}{\kappa^2 + 1}[\bm{E \cdot p} - (\bm{E:pp})\bm{p}], \label{spher:orient}
\end{align}
where $\bm{E}$ and $\bm{\Omega}$ are the rate-of-strain and vorticity tensors of the ambient planar linear flow. The canonical planar linear flows form a one-parameter family, the parameter being $\hat{\alpha}$, and Fig.\ref{fig:s4}a therefore organizes the spheroid trajectory topologies on the $\kappa-\hat{\alpha}$ plane with $\hat{\alpha}\in [-1,1]$. The particular eccentric linear flow sequence chosen in Fig.\ref{fig:s4}b has $\epsilon = -2$, $\phi_\omega = 0$, with $\theta_\omega$ varying from $50.74^\circ$ to $\frac{\pi}{2}$\,(accordingly, $1/\alpha$ varies from $0$ to $2\sqrt{2}$); the spheroid trajectory topologies in this case are organized on the $\kappa-\frac{(1-(\alpha/2))}{(1+(\alpha/2))}$ plane, the ratio $\frac{(1-(\alpha/2))}{(1+(\alpha/2))}$ being the analog of $\hat{\alpha}$ in the canonical case, and varying from $-1$ to $\approx -0.18$.
As evident from the two cases\,(for $\kappa = 100$ and $\infty$) shown in Fig.\ref{fig:s4}a, the sequence of unit-sphere trajectory topologies with increasing $\hat{\alpha}$ remains the same, regardless of $\kappa$, for the canonical case. The unit-sphere is foliated by closed orbits below the threshold curve, $\hat{\alpha} = \kappa^{-2}$, with these orbits corresponding to tumbling dynamics of the microstructure. Above the threshold curve, unit spheres are characterized by an open-trajectory topology \,(organized by a network of six fixed points comprising diametrically opposite pairs of stable nodes, unstable nodes and saddle points) that leads to flow-aligning dynamics. The threshold curve itself always corresponds to a unit sphere with a structurally unstable meridional trajectory topology: meridional trajectories running from one pole to the other in either direction are separated by a great circle of fixed points. For any $(\hat{\alpha},\kappa)$, the unit-sphere trajectories in Fig \ref{fig:s4}a correspond to the surface streamlines of a canonical planar linear flow, with this linear flow being the same as the ambient one for $\kappa = \infty$. Thus, the closed orbits that exist for $\hat{\alpha} < \kappa^{-2}$ correspond to the surface streamlines of an elliptic linear flow flow. For $\kappa = \infty$, this elliptic linear flow is the ambient one, and as mentioned in the manuscript, the trajectories shown are therefore spherical ellipses parameterized by the orbit constant $C = \left(\tan \theta \left( \gamma^2 \cos^2 \phi + \sin^2 \phi \right)^{1/2}\right) / \gamma $, where $\gamma = 1/\sqrt{-\hat{\alpha}}$. For any finite $\kappa$, the closed orbits are still spherical ellipses, with $1/\sqrt{-\hat{\alpha}}$ in the expression for $C$ being replaced by $\sqrt{\frac{\kappa^2 - \hat{\alpha}}{1 -  \kappa^2 \hat{\alpha}}}$.

Figure \ref{fig:s4}b shows the unit-sphere trajectory topologies, for prolate spheroids with $\kappa = 100$ and $\infty$, along an eccentric planar linear flow sequence; the parameters of this sequence are defined in the caption. The unit-sphere trajectory topologies shown correspond to values of $(1-(\alpha/2))/(1+(\alpha/2))$ below, on, and above the threshold curve\,(in blue). The latter curve, the analog of $\hat{\alpha} = \kappa^{-2}$ in the canonical case, may be inferred from the auxiliary linear equation associated with (\ref{spher:orient}), $\dot{\bm x} = \hat{\bm \Gamma} \bm{\cdot x}$, obtained by neglecting the final term in (\ref{spher:orient}). Here, $\hat{\bm \Gamma}= \bm{\Omega} + \frac{\kappa^2 - 1}{\kappa^2 + 1}\bm{E}$ and the threshold curve is obtained by equating the (cubic)\,discriminant $\Delta$ associated with $\hat{\bm \Gamma}$ to $0$. A comparison of Figs.\ref{fig:s4}a and b shows that there are crucial differences in the orientation dynamics relative to the canonical case. The orientation dynamics above the threshold curves remains qualitatively similar for the canonical and eccentric cases: six fixed points organize an open trajectory topology, leading to flow-aligning behavior. The dynamics on and below the threshold curve exhibits crucial differences. For the eccentric case, tumbling dynamics below the threshold curve, along generalized Jeffery orbits, only occurs for the infinitely slender spheroid\,($\kappa = \infty$) - these trajectories correspond to the surface streamlines of an eccentric elliptic flow. For any finite $\kappa$, the spheroid trajectories have a spiralling character, eventually converging towards a fixed orientation\,(the spiralling trajectories correspond to the surface streamlines of a $3D$ linear flow). Finally, the degenerate trajectory topologies along the threshold curve, for the eccentric case, differ qualitatively for finite $\kappa$ and $\kappa = \infty$: trajectories for $\kappa = \infty$ correspond to the surface streamlines of a degenerate parabolic flow, while those for any  finite $\kappa$ exhibit a degenerate node-saddle topology.